\documentclass[aps,prd,preprint,nofootinbib,11pt]{revtex4}
\usepackage{amsfonts}
\usepackage{mathrsfs}
\usepackage{graphicx}
\usepackage{amsmath}
\usepackage{amssymb}
\usepackage{subfigure}
\usepackage{epsfig}
\usepackage{graphicx}
\usepackage{color}
\newcommand{\bqa}{\begin{eqnarray}}
\newcommand{\eqa}{\end{eqnarray}}
\newcommand{\beq}{\begin{equation}}
\newcommand{\eeq}{\end{equation}}
\allowdisplaybreaks[1]
\graphicspath{{fig/}{dia/}} \DeclareGraphicsExtensions{.eps}
\begin{document}

\title{Spectrum of $J^{PC} = 0^{\pm\pm}$ Gluonic Hidden-Charm Tetraquark States \\[0.7cm]}

\author{Bing-Dong Wan$^{1,2}$\footnote{wanbd@lnnu.edu.cn}, Ming-Yang Yuan$^{1,2}$, Jun-Hao Zhang$^{1,2}$, and Yan Zhang$^{1,2}$ \vspace{+3pt}}

\affiliation{$^1$Department of Physics, Liaoning Normal University, Dalian 116029, China\\
$^2$ Center for Theoretical and Experimental High Energy Physics, Liaoning Normal University, Dalian 116029, China
}
\author{~\\~\\}

\begin{abstract}
\vspace{0.5cm}
We investigate gluonic hidden-charm tetraquark states composed of two valence quarks, two valence antiquarks and an explicit valence gluon. In the color configuration $[\bar{3}_c]_{c q}\otimes[8_c]_{G}\otimes[3_c]_{\bar{c}\bar{q}}$, a complete set of eight interpolating currents is constructed for states with quantum numbers $^{PC}=0^{++}$, $0^{-+},$ $0^{--}$, and $0^{+-}$. The corresponding mass spectra are systematically analysed within the QCD sum rule framework, including nonperturbative condensate contributions up to dimension eight. Our numerical analysis indicates the possible existence of six gluonic hidden-charm tetraquark states exhibiting stable behaviour in the adopted Borel windows. By replacing the charm quark with the bottom quark, masses for the corresponding hidden-bottom partners are also estimated. Possible production mechanisms and dominant decay channels are discussed, providing phenomenological guidance for experimental searches. These predicted states may be accessible at current and forthcoming facilities, including Belle II, PANDA, SuperB and LHCb, and thus offer an opportunity to probe explicit gluonic degrees of freedom in multiquark systems and deepen our understanding of nonperturbative QCD.
\end{abstract}
 \maketitle
\newpage
\section{Introduction}

Quantum Chromodynamics (QCD)~\cite{Gross:1973id,Politzer:1973fx,Wilson:1974sk} is established as the fundamental theory of the strong interaction. It is generally understood that the essential properties of hadrons are governed by the non-perturbative dynamics of QCD. While perturbation theory provides a robust and well-tested description at high energies, controlled and systematic methods for addressing non-perturbative effects remain limited. Consequently, elucidating the non-perturbative sector of QCD continues to represent a central challenge in modern high-energy physics. In this context, a variety of phenomenological and non-perturbative approaches have been developed and widely employed to investigate hadronic observables, including mass spectra, transition amplitudes, parton distribution functions, and fragmentation functions.

Within the framework of QCD and the quark model (QM)~\cite{GellMann:1964nj,Zweig}, hadronic configurations extending beyond the conventional mesons ($q\bar q$) and baryons ($qqq$) are in principle allowed. Such configurations, commonly referred to as unconventional or exotic hadronic states, include multiquark states, glueballs, and hybrid mesons. The investigation of these novel structures not only enlarges the hadronic spectrum but also provides a unique window into the underlying dynamics of QCD, particularly in its non-perturbative regime.

Since the beginning of the 21st century, remarkable progress in experimental facilities and analysis techniques in high-energy physics has led to the observation of a growing number of unconventional hadronic states, exemplified by the so-called XYZ states~\cite{Choi:2003ue,Aubert:2005rm,Belle:2011aa,Ablikim:2013mio,Liu:2013dau}. To date, more than forty such states or candidates have been reported, evoking a situation reminiscent of the “particle zoo” era of the last century. It is widely anticipated that many more novel hadronic structures will be discovered in the near future, signaling a renewed era of exploration in hadron spectroscopy. Elucidating the internal structure and dynamical origin of these newly observed states therefore constitutes one of the most compelling and timely challenges in contemporary hadron physics.

Motivated by the experimental successes in the exploration of the $XYZ$ and $P_c$ states, systematic searches for hybrid mesons in the charmonium and bottomonium sectors have been proposed and actively pursued~\cite{Olsen:2009gi,Olsen:2009ys,Godfrey:2008nc,Close:2007ny}. In general, a hybrid state is characterized by a constituent quark–antiquark pair accompanied by an explicit excitation of the gluonic degree of freedom. The identification and study of such states have become key objectives of several major experimental programs, including BESIII, GlueX, PANDA, and LHCb.

Although hybrid mesons have not yet been unambiguously established experimentally, a number of promising candidates have been reported in recent years as well as in earlier experiments. In particular, the BESIII Collaboration has recently observed a structure with quantum numbers $J^{PC}=1^{-+}$ at a mass around 1.855 GeV, denoted as $\eta_1(1855)$, in the $\eta\eta^{\prime}$ invariant mass distribution, with a statistical significance of $19\sigma$~\cite{BESIII:2022riz,BESIII:2022iwi}. Furthermore, three additional experimental candidates carrying the exotic quantum numbers $I^G J^{PC} = 1^-1^{-+}$ have been reported in the past, namely $\pi_1(1400)$~\cite{IHEP-Brussels-LosAlamos-AnnecyLAPP:1988iqi}, $\pi_1(1600)$~\cite{E852:2001ikk,COMPASS:2009xrl} and $\pi_1(2015)$~\cite{E852:2004gpn}.

Compared with conventional hybrid configurations, a novel class of exotic hybrid states composed of two valence quarks, two valence antiquarks, and a valence gluon has been proposed as a possible interpretation of some unusual features associated with the $X(6900)$ structure~\cite{Wan:2020fsk}. 

The narrow structure $X(6900)$ was first reported by the LHCb Collaboration in 2020 as a pronounced enhancement in the di-$J/\psi$ invariant mass spectrum around $6.9~\text{GeV}$. This observation was achieved with a statistical significance exceeding $5\,\sigma$ in proton--proton collisions at center-of-mass energies of $\sqrt{s} = 7$, 8, and $13~\text{TeV}$~\cite{Aaij:2020fnh}. It represented the first clear experimental evidence for a distinct structure in the $J/\psi$-pair mass spectrum, thereby opening a new window for exploring multiquark dynamics and exotic hadronic configurations.

In 2023, the existence of the $X(6900)$ structure was further corroborated by the CMS Collaboration, which also reported two additional enhancements, denoted as $X(6600)$ and $X(7100)$~\cite{CMS:2023owd}. Notably, the mass of $X(6900)$ lies approximately $700~\text{MeV}$ above the double-$J/\psi$ threshold, which is significantly larger than the typical mass splittings observed between ground and excited states in conventional quarkonium systems. Within the tetraquark hybrid framework, such a large energy separation can be naturally attributed to the presence of an excited gluonic degree of freedom. In this context, the emergence of a higher-mass structure such as $X(7100)$ is also anticipated.

In analogy with the proposed tetracharm hybrid configuration, one may also construct a class of gluonic charmonium-like states in which a charm--anticharm pair is partially replaced by a light quark pair, e.g., an up quark and an anti-down quark. In this scenario, the diquark pairs $cq$ and $\bar{c}\bar{q}$ are arranged in the color $\bar{\mathbf{3}}_c$ and $\mathbf{3}_c$ representations, respectively, within the $\mathrm{SU}(3)_c$ gauge group. Together with a dynamical gluon in the color octet representation, the resulting configuration of such gluonic charmonium-like states can be expressed as
\begin{equation}
[\bar{\mathbf{3}}_c]_{c q} \otimes [\mathbf{8}_c]_G \otimes [\mathbf{3}_c]_{\bar{c} \bar{q}} \, .
\end{equation}

In our previous work~\cite{Wan:2024pet}, tetraquark hybrid states with quantum numbers $J^{P}=0^{-}$, $1^{-}$, $0^{+}$, $1^{+}$ were studied. Owing to the increased complexity of the interpolating currents and the associated computational difficulties, the $C$-parity was not considered. 
Compared to our previous study in Ref.~\cite{Wan:2024pet}, where the interpolating currents do not possess definite charge-conjugation quantum numbers, the present work explicitly constructs currents with well-defined $C$-parity. As a result, the corresponding correlation functions in Ref.~\cite{Wan:2024pet} may receive contributions from mixtures of states with different $C$-parity, while in the present analysis we are able to project onto states with definite $J^{PC}=0^{\pm\pm}$.
This improvement is not merely technical. The use of $C$-parity eigenstates allows for a clearer classification of the hadronic states and leads to distinct structures in the operator product expansion. Therefore, the present work probes a different set of physical eigenstates and provides a complementary perspective on gluonic hidden-charm tetraquark systems.

In this paper, we investigate the $0^{\pm\pm}$ gluonic charmonium-like states within the framework of QCD sum rules (QCDSR)~\cite{Shifman}. 
Unlike phenomenological models, QCDSR is a first-principle, QCD-based nonperturbative approach that systematically incorporates vacuum condensates through the operator product expansion, thereby providing a self-consistent description of hadronic properties governed by nonperturbative QCD dynamics. 
Over the past decades, this method has achieved remarkable success in the study of hadron spectroscopy and exotic states~\cite{Shifman,Albuquerque:2013ija,Wang:2013vex,Govaerts:1984hc,Reinders:1984sr,P.Col,Narison:1989aq,Tang:2021zti,Qiao:2014vva,Qiao:2015iea,Tang:2019nwv,Wan:2019ake,Wan:2020oxt,Wan:2021vny,Wan:2022xkx,Zhang:2022obn,Wan:2022uie,Wan:2023epq,Wan:2024dmi,Tang:2024zvf,Li:2024ctd,Zhao:2023imq,Yin:2021cbb,Yang:2020wkh,Wan:2024cpc,Wan:2024pet,Wan:2024ykm,Wan:2025xhf,Wan:2025fyj,Wan:2025zau,Wan:2025bdr,Wan:2025sae,Wang:2021qmn,Agaev:2022pis,Agaev:2025llz,Barsbay:2025vjq,Wang:2025sic,Wang:2020cme}. Beyond mass spectra, it has been successfully extended to investigate a variety of other dynamical and structural properties of exotic hadrons. For instance, three-point and light-cone sum rules have been widely employed to calculate hadronic coupling constants and strong decay widths \cite{Wang:2019bbl,Dias:2013xfa,Huang:1998rq}, as well as to probe the internal charge distribution of multiquark states through the evaluation of their electromagnetic multipole moments \cite{Wang:2017dce,Ozdem:2018qeh}. By investigating these non-mass observables, the sum rule method provides crucial complementary insights into the phenomenological landscape of exotic hadron spectroscopy.

To formulate the QCD sum rules, the first step is to construct appropriate interpolating currents that carry the same quantum numbers as the hadron of interest. Using these currents, one then defines the two-point correlation function, which admits two distinct representations: a QCD (theoretical) side, evaluated via the operator product expansion, and a phenomenological (hadronic) side, expressed in terms of physical intermediate states. By matching these two representations through a dispersion relation and applying a Borel transformation, the sum rule is established, from which the hadron mass can be extracted in a systematic manner.

The rest of this paper is organized as follows. 
In Sec.~\ref{formalism}, we present a brief overview of the QCD sum rule method together with the essential formulas used in our analysis. 
The numerical results and discussions are given in Sec.~\ref{numerical}. 
Possible production mechanisms and decay modes of the hybrid states are discussed in Sec.~\ref{decay}. 
Finally, a brief summary is presented in the last section.

\section{Formalism}\label{formalism}
In the QCD sum rule framework, the two-point correlation function is defined as
\begin{eqnarray}
\Pi(q^2) &=& i \int d^4 x e^{i q \cdot x} \langle 0 | T \{ j (x),\;  j^\dagger (0) \} |0 \rangle\;.\label{two-points-a}
\end{eqnarray}

To evaluate the mass spectrum of gluonic tetraquark states, the corresponding interpolating currents are constructed~\cite{Wan:2020fsk}. For the $0^{++}$ state, the currents take the forms:
\begin{eqnarray}
j^{A}_{0^{++}} (x)&=& g_s \epsilon_{ikl}\epsilon_{jmn} [q_k^T C \gamma_\mu c_l]\frac{\lambda_{ij}^a}{2} G_{\mu\nu}^a [\bar{q}_m \gamma_\nu C \bar{c}_n^{T}]\, , \label{current-0++A} \\
j^{B}_{0^{++}}(x) &=& g_s \epsilon_{ikl}\epsilon_{jmn} [q_k^T C \gamma_\mu\gamma_5 c_l]\frac{\lambda_{ij}^a}{2} G_{\mu\nu}^a [\bar{q}_m \gamma_\nu\gamma_5 C \bar{c}_n^{T}]\, , \label{current-0++B} \\
j^{C}_{0^{++}} (x)&=& \frac{g_s \epsilon_{ikl}\epsilon_{jmn}}{\sqrt{2}}{\{}[q_k^T C \gamma_\mu c_l]\frac{\lambda_{ij}^a}{2} \tilde{G}_{\mu\nu}^a [\bar{c}_m \gamma_\nu \gamma_5 C \bar{q}_n^{T}]\,\nonumber\\
&+& [q_k^T C \gamma_\mu \gamma_5 c_l]\frac{\lambda_{ij}^a}{2} \tilde{G}_{\mu\nu}^a [\bar{c}_m \gamma_\nu C \bar{q}_n^{T}]{\}}\,. \label{current-0++C} 
\end{eqnarray}
Here, $q$ denotes the up or down quarks; $G_{\mu\nu}^a$ and $\tilde{G}{\mu\nu}^a$ represent the gluon field strength tensor and its dual, with $\tilde{G}{\mu\nu}^a = \frac{1}{2}\epsilon_{\mu\nu\alpha\beta} G^{a,\;\alpha\beta}$; $g_s$ is the QCD coupling constant; $\lambda^a$ are the Gell-Mann matrices; $C$ denotes the charge conjugation matrix; and $i$, $j$, $k$, $\dots$ are color indices. 

For the $0^{-+}$ state, the interpolating currents are:
\begin{eqnarray}
j^{A}_{0^{-+}} (x)&=& g_s \epsilon_{ikl}\epsilon_{jmn} [q_k^T C \gamma_\mu c_l]\frac{\lambda_{ij}^a}{2} \tilde{G}_{\mu\nu}^a [\bar{q}_m \gamma_\nu C \bar{c}_n^{T}]\, , \label{current-0-+A} \\
j^{B}_{0^{-+}}(x) &=& g_s \epsilon_{ikl}\epsilon_{jmn} [q_k^T C \gamma_\mu\gamma_5 c_l]\frac{\lambda_{ij}^a}{2} \tilde{G}_{\mu\nu}^a [\bar{q}_m \gamma_\nu\gamma_5 C \bar{c}_n^{T}]\, , \label{current-0-+B}\\
j^{C}_{0^{-+}} (x)&=& \frac{g_s \epsilon_{ikl}\epsilon_{jmn}}{\sqrt{2}}{\{}[q_k^T C \gamma_\mu c_l]\frac{\lambda_{ij}^a}{2} G_{\mu\nu}^a [\bar{q}_m \gamma_\nu \gamma_5 C \bar{c}_n^{T}]\, \nonumber\\
&+&[q_k^T C \gamma_\mu \gamma_5 c_l]\frac{\lambda_{ij}^a}{2} G_{\mu\nu}^a [\bar{q}_m \gamma_\nu C \bar{c}_n^{T}]{\}}\,. \label{current-0-+C} 
\end{eqnarray}

For the $0^{--}$ and $0^{+-}$ states, the currents are:
\begin{eqnarray}
j^{A}_{0^{--}} (x)&=&\frac{ g_s \epsilon_{ikl}\epsilon_{jmn}{\{}}{\sqrt{2}}[q_k^T C \gamma_\mu c_l]\frac{\lambda_{ij}^a}{2} G_{\mu\nu}^a [\bar{q}_m \gamma_\nu \gamma_5 C \bar{c}_n^{T}]\, \nonumber\\
&-&[q_k^T C \gamma_\mu c_l]\frac{\lambda_{ij}^a}{2} G_{\mu\nu}^a [\bar{q}_m \gamma_\nu \gamma_5 C \bar{c}_n^{T}]{\}}\,,\\ \label{current-0--A} 
j^{A}_{0^{+-}} (x)&=& \frac{g_s \epsilon_{ikl}\epsilon_{jmn}{\{}}{\sqrt{2}}[q_k^T C \gamma_\mu c_l]\frac{\lambda_{ij}^a}{2} \tilde{G}_{\mu\nu}^a [\bar{c}_m \gamma_\nu \gamma_5 C \bar{q}_n^{T}]\,\nonumber\\
&-& [q_k^T C \gamma_\mu \gamma_5 c_l]\frac{\lambda_{ij}^a}{2} \tilde{G}_{\mu\nu}^a [\bar{c}_m \gamma_\nu C \bar{q}_n^{T}]{\}}\,. \label{current-0+-A} 
\end{eqnarray}

On the OPE side, the correlation function can be expressed as
\begin{eqnarray}
\Pi^{OPE}(q^2) &=& \int_{s_{min}}^{\infty} d s
\frac{\rho^{OPE} (s)}{s - q^2} \; ,
\label{OPE-hadron}
\end{eqnarray}
where $\rho^{\rm OPE}(s) = \mathrm{Im}[\Pi^{\rm OPE}(s)]/\pi$ is the spectral density, and $s_{\rm min} = (2 m_c + 2 m_q)^2$. Up to dimension-eight condensates, it reads
\begin{eqnarray}
\rho^{OPE}(s)  =  \rho^{pert}(s) + \rho^{\langle \bar{q} q
\rangle}(s) +\rho^{\langle G^2 \rangle}(s) + \rho^{\langle \bar{q} G q \rangle}(s)
+ \rho^{\langle \bar{q} q \rangle^2}(s) + \rho^{\langle G^3 \rangle}(s) 
+ \rho^{\langle \bar{q} q \rangle\langle \bar{q} G q \rangle}(s)  . \label{rho-OPE}
\end{eqnarray}

\begin{figure}
\includegraphics[width=12cm]{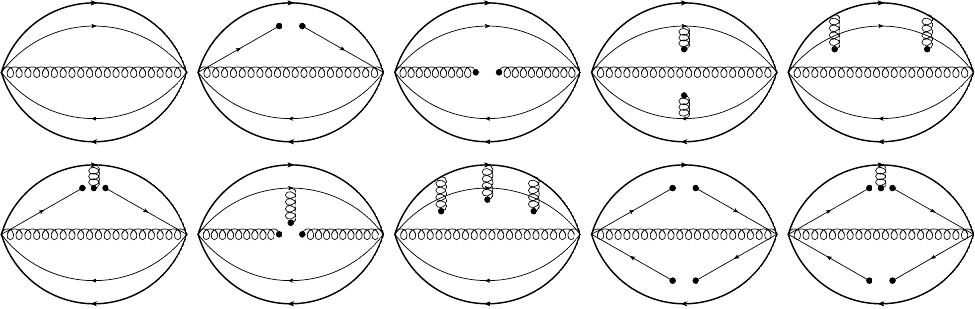}
\caption{The typical Feynman diagrams associated with the correlation function are shown, where the thick solid line represents the heavy quark, the thin solid line denotes the light quark, and the wavy line corresponds to the gluon.} \label{figfeyn}
\end{figure}

The Feynman diagrams corresponding to these contributions are illustrated in Fig.~\ref{figfeyn}, where the thick solid line represents the heavy quark, the thin solid line denotes the light quark, and the wavy line corresponds to the gluon. Each diagram provides a visual representation of how different condensates and perturbative contributions enter the correlation function, with lower-dimensional condensates typically associated with simpler topologies and higher-dimensional condensates corresponding to more complex gluon insertions or multi-quark interactions.

On the phenomenological side, the correlation function is written as
 \begin{eqnarray}
\Pi^{phen}(q^2) & = & \frac{\lambda^2}{m^2 - q^2} + \frac{1}{\pi} \int_{s_0}^\infty d s \frac{\rho(s)}{s - q^2} \; , \label{hadron}
\end{eqnarray}
where $m$, $\lambda$, $\rho(s)$, and $s_0$ denote the hadron mass, coupling constant, spectral density of higher states, and continuum threshold.

To suppress contributions from higher-dimensional condensates and continuum states, a Borel transformation is applied:
\begin{eqnarray}
\hat{B}[f(Q^2)]:=\lim_{Q^2\rightarrow \infty,n\rightarrow \infty
\atop{Q^2 / n}={M_B^2}}
\frac{(-Q^2)^n}{(n-1)!}\left(\frac{d}{dQ^2}\right)^n\;f(Q^2)\ .
\end{eqnarray}

After the Borel transformation and applying quark–hadron duality, the QCD sum rule reads
\begin{eqnarray}
\lambda^2 e^{-m^2/M_B^2} + \int_{s_0}^\infty d s \rho^{OPE}(s) e^{-s/M_B^2}= \int_{s_{min}}^\infty d s \rho^{OPE}(s) e^{-s/M_B^2}\;,
\end{eqnarray}
from which the tetraquark mass is extracted as
\begin{eqnarray}
m(s_0, M_B^2) &=& \sqrt{-\frac{L_1(s_0, M_B^2)}{L_0(s_0, M_B^2)}} \; , \label{mass-Eq}
\end{eqnarray}
with
\begin{eqnarray}
L_0(s_0, M_B^2) & = & \int_{s_{min}}^{s_0} d s \; \rho(s) e^{-s / M_B^2} \; , \label{L0}\\
L_1(s_0, M_B^2) & = & \frac{\partial}{\partial{\frac{1}{M_B^2}}}{L_0(s_0, M_B^2)} \; .
\end{eqnarray}

\section{Numerical analysis}\label{numerical}
To perform numerical evaluations, the following widely adopted input parameters are used~\cite{Shifman, Albuquerque:2013ija, Reinders:1984sr, P.Col, Narison:1989aq}:
\begin{align}
& m_c(m_c) = \overline{m}_c = (1.275 \pm 0.025)\; \text{GeV}, \quad
m_b(m_b) = \overline{m}_b = (4.18 \pm 0.03)\; \text{GeV}, \nonumber\\
& m_u = 2.16^{+0.49}_{-0.26} \;\text{MeV}, \quad
m_d = 4.67^{+0.48}_{-0.17}\; \text{MeV}, \nonumber\\
& \langle \bar{q} q \rangle = - (0.23 \pm 0.03)^3 \;\text{GeV}^3, \quad
\langle g_s^2 G^2 \rangle = 0.88 \;\text{GeV}^4, \quad
\langle g_s^3 G^3 \rangle = 0.045 \;\text{GeV}^6, \nonumber\\
& \langle \bar{q} g_s \sigma \cdot G q \rangle = m_0^2 \langle \bar{q} q \rangle, \quad
m_0^2 = (0.8 \pm 0.2) \;\text{GeV}^2 .
\end{align}

The $\overline{\rm MS}$ running masses are used for the heavy quarks. All quantities are evaluated at a typical renormalization scale $\mu = 2~\mathrm{GeV}$, which is of the order of the charm-quark mass.

The leading-order strong coupling constant is taken as
\begin{eqnarray}
\alpha_s(\mu^2) = \frac{4 \pi}{\left(11 - \frac{2}{3} n_f \right) \ln \left( \frac{\mu^2}{\Lambda_{\rm QCD}^2} \right)} \, ,
\end{eqnarray}
with $\Lambda_{\rm QCD} = 300$ MeV and $n_f$ being the number of active quark flavors. In the numerical analysis, the renormalization scale is chosen to be of the same order as the Borel parameter, i.e., $\mu^2 \sim M_B^2$.

We note that the interpolating currents considered in this work contain explicit gluonic field strength tensors and are not eigenstates of the renormalization group. A complete treatment would require the inclusion of anomalous dimensions and possible operator mixing effects. In the present work, we restrict ourselves to leading order in $\alpha_s$ and do not explicitly include these effects. The associated scale dependence is expected to be within the overall theoretical uncertainties of the sum rule analysis.
We have checked that varying the scale in the range $\mu = 2\text{--}3~\mathrm{GeV}$ does not lead to significant changes in the extracted masses.

In evaluating the higher-dimensional condensates, we adopt the vacuum saturation approximation, which allows us to factorize four-quark condensates into products of quark condensates.

It is known that for multiquark systems, violations of factorization may be sizable. To estimate this effect, we introduce a parameter $\kappa$ to parametrize deviations from vacuum saturation,
\[
\langle \bar{q} q \bar{q} q \rangle = \kappa \langle \bar{q} q \rangle^2.
\]
We vary $\kappa$ in the range $0.5 \sim 1.5$ and find that the extracted masses are relatively stable, with variations within the theoretical uncertainties of the present analysis.

The extracted masses of the tetraquark hybrid states depend on two auxiliary parameters, the Borel parameter $M_B^2$ and the continuum threshold $s_0$. Their values are constrained by standard QCD sum rule criteria to ensure the reliability of the analysis~\cite{Shifman, Reinders:1984sr, P.Col, Albuquerque:2013ija}.

The determination of the Borel window is based on a balance between the convergence of the operator product expansion (OPE) and the dominance of the pole contribution.

The lower bound of the Borel window is determined by requiring good convergence of the OPE. In practice, this is achieved by ensuring that the contribution from higher-dimensional condensates remains sufficiently small compared to the total OPE contribution. We quantify this using the ratio
\begin{eqnarray}
  R^{\rm OPE} = \left| \frac{L^{\rm dim=8}_{0}(s_0, M_B^2)}{L_{0}(s_0, M_B^2)}\right|\, .
\end{eqnarray}
A reliable working region requires this ratio to be sufficiently small, indicating good convergence of the truncated OPE.

The upper bound of the Borel window is constrained by the pole contribution (PC), which ensures that the extracted mass is dominated by the ground-state resonance rather than the continuum. The pole contribution is defined as
\begin{eqnarray}
  R^{\rm PC} = \frac{L_{0}(s_0, M_B^2)}{L_{0}(\infty, M_B^2)} \, .
\end{eqnarray}
In this work, we require the pole contribution to be larger than $40\%$, which is a commonly adopted criterion in QCD sum rule analyses of multiquark and hybrid systems. We have also checked that imposing a more stringent condition, such as $R^{\rm PC} > 50\%$, does not significantly affect the extracted masses within uncertainties.

Therefore, the Borel window $M_B^2 \in [M_{\min}^2, M_{\max}^2]$ is determined by requiring that both OPE convergence (lower bound) and pole dominance (upper bound) are simultaneously satisfied.

For each fixed value of $s_0$, we determine the corresponding Borel window and examine the stability of the extracted mass with respect to $M_B^2$. The optimal working region is identified as the one where a stable Borel plateau exists.

The continuum threshold $s_0$ is then varied within a physically motivated range, typically in steps of $0.2\;\text{GeV}$, to determine its optimal value and estimate the associated uncertainty~\cite{Qiao:2014vva, Qiao:2015iea}. The final results are obtained from the region where both the Borel stability and the above criteria are satisfied, and the uncertainties include the variation with respect to both $M_B^2$ and $s_0$.

\begin{figure}
\includegraphics[width=6.8cm]{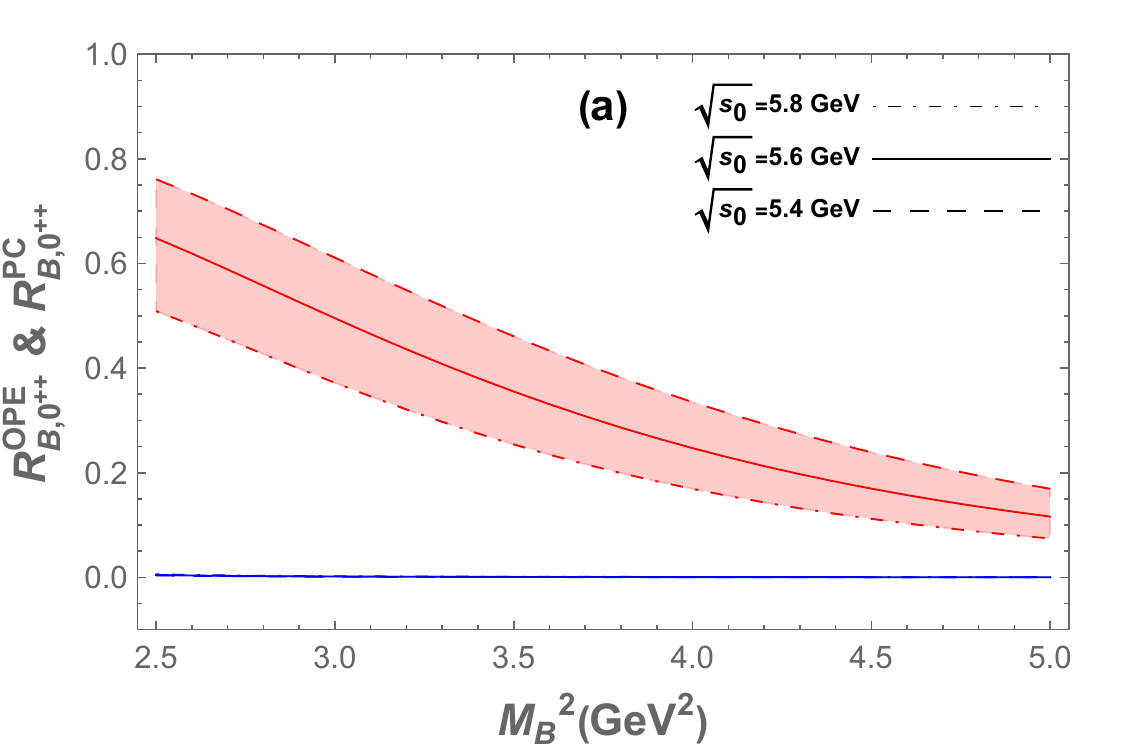}
\includegraphics[width=6.8cm]{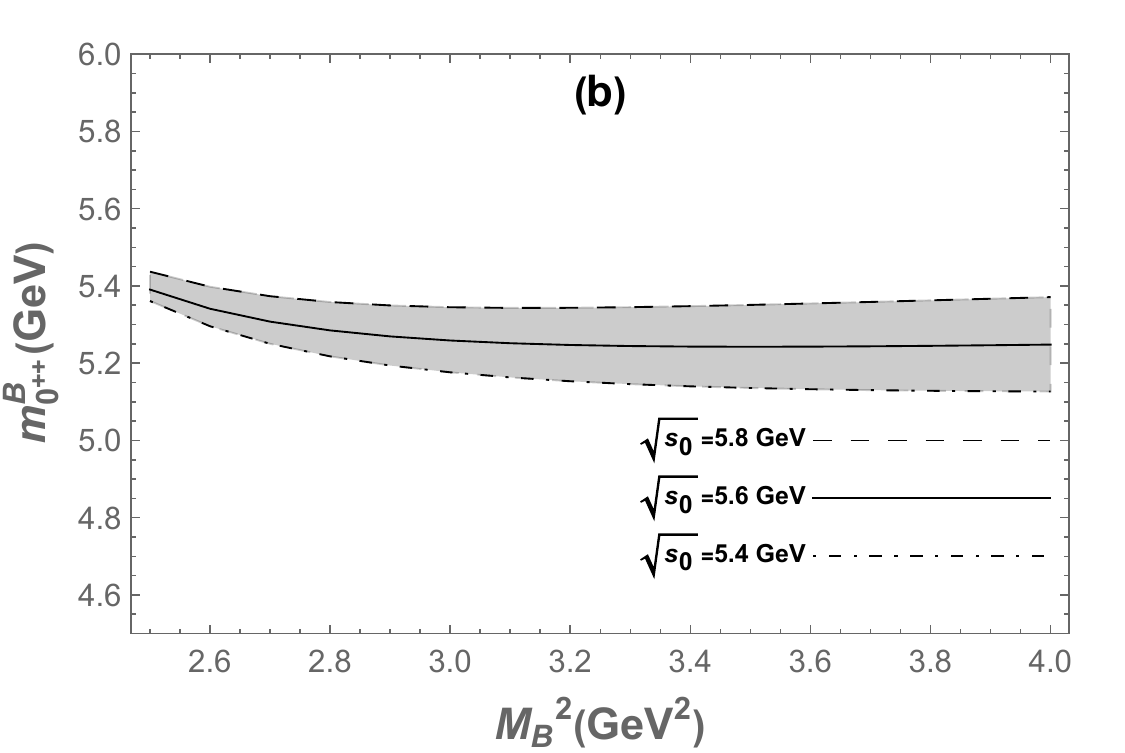}
\caption{ (a) The ratios ${R_{0^{++}}^{B,\;OPE}}$ and ${R_{0^{++}}^{B,\;PC}}$ as functions of the Borel parameter $M_B^2$ for different values of $\sqrt{s_0}$, where blue lines represent ${R_{0^{++}}^{B,\;OPE}}$ and red lines denote ${R_{0^{++}}^{B,\;PC}}$. (b) The mass $M_{0^{++}}^{B}$ as a function of the Borel parameter $M_B^2$ for different values of $\sqrt{s_0}$.} \label{fig0++B}
\end{figure}

With the above preparations, we are now in a position to numerically evaluate the mass spectrum of the tetraquark hybrid states. As an illustrative example, the ratios ${R_{0^{++}}^{B,\;\mathrm{OPE}}}$ and ${R_{0^{++}}^{B,\;\mathrm{PC}}}$ associated with the current in Eq.~(\ref{current-0++B}) are displayed in Fig.~\ref{fig0++B}(a) as functions of the Borel parameter $M_B^2$ for different values of the continuum threshold, namely $\sqrt{s_0} = 5.4$, $5.6$, and $5.8~\mathrm{GeV}$. The corresponding dependence of $m_{0^{++}}^{B}$ on $M_B^2$ is shown in Fig.~\ref{fig0++B}(b).
The optimal Borel window is determined to be $2.6 \leq M_B^2 \leq 3.4 \; \mathrm{GeV}^2$
within which a smooth and stable plateau emerges in the $m_{0^{++}}^{B}$--$M_B^2$ curve. This behavior signals the reliability of the extracted mass for the possible $0^{++}$ tetraquark hybrid state. Accordingly, the mass $m_{0^{++}}^{B}$ is obtained as
\begin{eqnarray}
&&m_{0^{++}}^{B} = (5.24 \pm 0.10) \, \text{GeV}  \; ,\label{eq-mass-0++B}\\
&&\lambda_{0^{++}}^{B}=(1.32 \pm 0.40)\times 10^{-3}\,\text{GeV}^5\,.
\end{eqnarray}

Following the same procedure, the OPE convergence, pole contribution, and the dependence of the extracted masses on the Borel parameter $M_B^2$ are presented in Appendix~\ref{App_B}. The resulting mass values are then obtained and collected in Table~\ref{mass_c}.
For the currents in Eqs.~(\ref{current-0++A}) and (\ref{current-0-+A}), however, no appropriate Borel window can be identified in which the masses exhibit sufficient stability with respect to variations of $M_B^2$. This suggests that these currents have no strong coupling to hidden-charm tetraquark hybrid states.

\begin{table}
\begin{center}
\renewcommand\tabcolsep{10pt}
\caption{The continuum thresholds, Borel parameters, and predicted masses of the hidden-charm tetraquark hybrid states.}\label{mass_c}
\begin{tabular}{cccccc}\hline\hline
$J^P$             &Current   & $\sqrt{s_0}\;(\text{GeV})$     &$M_B^2\;(\text{GeV}^2)$ &$M^X\;(\text{GeV})$  & $\lambda^X\;(10^{-3}\,\text{GeV}^5)$      \\ \hline
$0^{++}$        &$A$         & $-$                                        & $-$                                   &$-$                            &$-$  \\
                      &$B$         & $5.6\pm0.2$                         &$2.6-3.4$                          &$5.24\pm0.10$          &$1.32 \pm 0.40$ \\
                      &$C$         & $5.7\pm0.2$                         &$2.6-3.3$                          &$5.33\pm0.10$          &$1.12 \pm 0.31$\\\hline
$0^{-+}$         &$A$         & $-$                                        &$-$                                    &$-$                             &$-$\\
                      &$B$         & $5.7\pm0.2$                         &$2.6-3.5$                          &$5.32\pm0.12$.          &$1.48 \pm 0.46$\\
                      &$C$         & $5.9\pm0.2$                         &$2.6-3.7$                          &$5.46\pm0.12$           &$1.54 \pm 0.53$\\\hline
$0^{--}$         &$A$         & $5.8\pm0.2$                          &$2.6-3.9$                          &$5.40\pm0.13$           &$1.53 \pm 0.59$     \\
$0^{+-}$         &$A$         & $5.6\pm0.2$                         &$2.6-3.5$                          &$5.28\pm0.12$         &$1.11 \pm 0.37$ \\
  \hline
 \hline
\end{tabular}
\end{center}
\end{table}

For the bottom sector, performing the same procedure but with $m_c$ replaced by $m_b$, readers can find the OPE, pole contribution and the masses as functions of Borel parameter $M^2_B$ in the Appendix \ref{App_B}, we can easily obtain the masses of the hidden-bottom tetraquark hybrid states. The relevant numerical results are tabulated in Table~\ref{mass_b}.

In the present work, the QCD sum rule analysis is performed at leading order in $\alpha_s$, which is the standard approximation in many studies of multiquark and hybrid systems. Higher-order perturbative corrections are expected to provide quantitative modifications, but are unlikely to change the qualitative features of the results.
For the heavy quarks, we adopt the $\overline{\rm MS}$ running masses, which are commonly used in QCD sum rule analyses and exhibit better convergence behavior than the pole mass scheme. This choice is consistent with the scale-dependent treatment of the operator product expansion.
Possible uncertainties associated with the truncation of the perturbative expansion and the choice of mass scheme are considered to be part of the overall theoretical uncertainties.
We expect that NLO corrections mainly lead to moderate shifts in the numerical results without altering the qualitative conclusions.

\begin{table}
\begin{center}
\renewcommand\tabcolsep{10pt}
\caption{The continuum thresholds, Borel parameters, and predicted masses of the hidden-bottom tetraquark hybrid states.}\label{mass_b}
\begin{tabular}{cccccc}\hline\hline
$J^P$             &Current   & $\sqrt{s_0}\;(\text{GeV})$     &$M_B^2\;(\text{GeV}^2)$ &$M^X\;(\text{GeV})$ & $\lambda^X\;(10^{-5}\,\text{GeV}^5)$      \\ \hline
$0^{++}$       &$A$         & $-$                                        & $-$                                   &$-$                            &$-$\\
                      &$B$         & $11.6\pm0.2$                        &$6.3-7.6$                          &$11.15\pm0.11$        &$7.45 \pm 4.45$  \\
                      &$C$         & $11.7\pm0.2$                       &$6.2-7.4$                          &$11.23\pm0.11$         &$5.46 \pm 3.18$\\\hline
$0^{-+}$        &$A$         & $-$                                        &$-$                                    &$-$                            &$-$ \\
                      &$B$         & $11.7\pm0.2$                        &$6.5-7.9$                          &$11.22\pm0.11$.       &$9.81 \pm 5.94$  \\
                      &$C$         & $11.8\pm0.2$                       &$6.5-7.7$                          &$11.29\pm0.11$         &$7.51 \pm 4.20$  \\\hline
$0^{--}$         &$A$         & $11.7\pm0.2$                        &$6.3-7.4$                          &$11.23\pm0.11$         &$5.59 \pm 3.07$           \\
$0^{+-}$        &$A$         & $11.6\pm0.2$                       &$6.2-7.2$                          &$11.17\pm0.11$          &$4.49 \pm 2.38$\\
 \hline
 \hline
\end{tabular}
\end{center}
\end{table}

The uncertainties quoted in Tables~\ref{mass_c} and~\ref{mass_b} are mainly induced by the variations of the input parameters, including the quark masses, the condensates, and the continuum threshold $\sqrt{s_0}$.

\section{Production and decay mechanisms of the $0^{\pm\pm}$ tetraquark--hybrid states}\label{decay}

According to our QCD sum rule analysis, the masses of the hidden-charm tetraquark--hybrid states lie in the range $5.14~\mathrm{GeV} \le m_{X} \le 5.58~\mathrm{GeV}$. This mass region is above the open-charm thresholds and overlaps with the energy domain where a variety of exotic charmonium-like structures, commonly referred to as the XYZ states, have been reported. Consequently, the production and decay mechanisms of such tetraquark--hybrid states are expected to be rich and experimentally accessible.

\subsection{Production mechanisms}

\subsubsection{Gluon-dominated production in hadron colliders}

Due to the explicit valence gluon in their internal structure, tetraquark--hybrid states can be efficiently produced in gluon-rich environments, such as high-energy proton--proton collisions at the LHC. Typical partonic subprocesses include
\begin{equation}
g + g \rightarrow c \bar{c} + q \bar{q} + g,
\end{equation}
where the additional gluon becomes an intrinsic constituent of the hybrid configuration. Non-perturbative hadronization of such partonic systems may lead to the formation of compact tetraquark--hybrid states, making LHC and LHCb particularly suitable for searches in invariant mass spectra of charmonium plus light hadrons.

\subsubsection{Production in $e^{+}e^{-}$ annihilation}

In the energy region around 5--6~GeV, production via $e^{+}e^{-}$ annihilation is influenced by intermediate vector charmonium resonances, e.g.,
\begin{equation}
e^{+}e^{-} \rightarrow \psi^{*} \rightarrow X_{\rm hybrid} + \pi (\pi / \eta),
\end{equation}
where $X_{\rm hybrid}$ denotes a hidden-charm tetraquark--hybrid state. Such processes are within the capabilities of Belle II, and SuperKEKB.

\subsubsection{Production in $B$-meson decays}

Weak decays of $B$ mesons also provide a production mechanism:
\begin{equation}
B \rightarrow X_{\rm hybrid} + K (\pi),
\end{equation}
where the underlying quark-level transition $b \to c \bar{c} s$ naturally produces a hidden-charm pair, and an excited gluon may become part of the hybrid configuration.

\subsection{Decay mechanisms of $0^{\pm\pm}$ states}

The decay patterns of tetraquark--hybrid states are governed by their spin-parity-charge conjugation quantum numbers ($J^{PC}$), mass thresholds, and internal gluonic content. The dominant decay modes can be summarized in Table~\ref{tab:decay}.

\begin{table}[htbp]
\centering
\caption{Dominant decay channels of the $0^{\pm\pm}$ hidden-charm tetraquark--hybrid states. Here, "OC" stands for open-charm meson decays, "HC" for hidden-charm meson decays, "Rad" for radiative decays, and "BB" for baryon--antibaryon decays.}
\label{tab:decay}
\begin{tabular}{c l l}
\hline\hline
$J^{PC}$ & Decay channels & Type \\
\hline
$0^{++}$ & $D \bar{D},\ D^* \bar{D}^*,\ D_s \bar{D}_s$ & OC \\
         & $J/\psi\, \omega,\ \eta_c\, f_0(980),\ \chi_{c0}\, \eta$ & HC \\
         & $J/\psi\, \gamma,\ \chi_{c1}\, \gamma$ & Rad \\
         & $\Lambda_c \bar{\Lambda}_c,\ \Sigma_c \bar{\Sigma}_c,\ \Xi_c \bar{\Xi}_c$ & BB \\
\hline
$0^{--}$ & $J/\psi\, \pi^0,\ J/\psi\, \eta,\ D \bar{D}^* (L=1)$ & HC/OC \\
         & $J/\psi + \pi^+ \pi^-,\ D \bar{D} + \pi$ & Multi-body \\
         & $\Lambda_c \bar{\Lambda}_c,\ \Sigma_c \bar{\Sigma}_c,\ \Xi_c \bar{\Xi}_c$ & BB \\
\hline
$0^{+-}$ & $J/\psi + \pi + \pi,\ \eta_c + \rho + \pi$ & HC/Multi-body \\
         & $D \bar{D} + \pi,\ D^* \bar{D} + \pi$ & OC/Multi-body \\
         & $\Lambda_c \bar{\Lambda}_c,\ \Sigma_c \bar{\Sigma}_c,\ \Xi_c \bar{\Xi}_c$ & BB \\
\hline
$0^{-+}$ & $D \bar{D}^*,\ D^* \bar{D}^* (L=1)$ & OC \\
         & $J/\psi\, \eta,\ \eta_c\, \omega$ & HC \\
         & $J/\psi\, \gamma,\ \eta_c\, \gamma$ & Rad \\
         & $\Lambda_c \bar{\Lambda}_c,\ \Sigma_c \bar{\Sigma}_c,\ \Xi_c \bar{\Xi}_c$ & BB \\
\hline\hline
\end{tabular}
\end{table}

It should be noted that the masses of the tetraquark--hybrid states in our calculation are close in magnitude to those of the hidden-charm hexaquarks reported in Refs.~\cite{Wan:2019ake,Wang:2021qmn,Liu:2021gva}. The main difference lies in the decay branching ratios into $\Xi_c \bar{\Xi}_c$ final states. For tetraquark--hybrid states, the branching ratios of $\Xi_c \bar{\Xi}_c$ are comparable to those of $\Sigma_c \bar{\Sigma}_c$, while for hidden-charm hexaquarks, the Cabibbo-Kobayashi-Maskawa suppression renders the $\Xi_c \bar{\Xi}_c$ branching ratios relatively small. This distinction may serve as a phenomenological handle to discriminate between these two types of exotic states in experimental searches.

The multi-body and radiative decay channels, particularly for the exotic $0^{--}$ and $0^{+-}$ states, offer distinctive signatures for experimental investigations. Analyses of invariant mass spectra in $D^{(*)}\bar{D}^{(*)}$, $J/\psi \pi\pi$, and $J/\psi \gamma$ channels at Belle II, SuperKEKB, PANDA, and LHCb are expected to be crucial in identifying these tetraquark--hybrid states and probing the role of gluonic excitations in hadron spectroscopy.

\section{Summary}

In summary, we have studied the hidden-charm and hidden-bottom tetraquark hybrid states with quantum numbers $J^{PC}=0^{\pm\pm}$, composed of two valence quarks, two valence antiquarks, and an explicit valence gluon. A complete set of eight interpolating currents was constructed with the color configuration
\begin{equation}
[\bar{3}_c]_{cq} \otimes [8_c]_G \otimes [3_c]_{\bar c \bar q}.
\end{equation}
Within the framework of QCD sum rules, we systematically evaluated the corresponding mass spectra. Our numerical analysis suggests the possible existence of six hidden-charm and hidden-bottom tetraquark hybrid states, and the resulting mass predictions are summarized in Tabs.~\ref{mass_c} and \ref{mass_b} for the charm and bottom sectors, respectively.

It should be noted that the interpolating currents used in this work are not unique and may mix with other operators carrying the same quantum numbers, such as hybrid-type $\bar{Q}QG$ operators or derivative-type four-quark currents. A complete analysis would require a correlation-matrix approach including multiple operators.
In the present work, we adopt the standard single-current approximation in QCD sum rules. Within this framework, the extracted masses correspond to states that can be coupled to the chosen currents, and should not be interpreted as strictly independent physical eigenstates.

It should be noted that, via Fierz rearrangement, the interpolating currents used in this work may couple not only to compact tetraquark configurations but also to two-meson scattering states with the same quantum numbers.
In the QCD sum rule framework, it is generally difficult to completely separate compact states from scattering contributions. The extracted masses are obtained in the Borel window where the pole contribution is reasonably large and the OPE converges well. The observed stability of the sum rules supports the interpretation that the signals are dominated by low-lying resonant contributions.
However, possible mixing with scattering states cannot be completely excluded, and the results should be interpreted with this caveat.

We further investigated the possible production and decay channels of the hidden-charm tetraquark hybrid states. Since the parent particles involved in these processes can be copiously produced in current experiments, the corresponding tetraquark hybrid states may be accessible at facilities such as BELLE~II, PANDA, Super-B, and LHCb.

Special attention has been paid to the states with exotic quantum numbers $J^{PC}=0^{--}$ and $0^{+-}$. While these quantum numbers cannot be realized by conventional $q\bar q$ mesons, they may arise from different nonconventional configurations, including three-gluon ($ggg$) glueballs. Therefore, an assignment based solely on quantum numbers is not sufficient to identify the underlying structure of such states.

In this work, the tetraquark hybrid interpretation can nevertheless be distinguished from the three-gluon glueball scenario in several aspects. The presence of heavy quarks implies a significantly higher mass scale for the hidden-charm and hidden-bottom tetraquark hybrid states compared to the expected masses of light $ggg$ glueballs. Moreover, the color configuration
\(
[\bar{3}_c]_{cq} \otimes [8_c]_G \otimes [3_c]_{\bar c \bar q}
\)
corresponds to a diquark--antidiquark core coupled to a valence gluon, leading to characteristic decay patterns involving heavy quarkonium states. In contrast, three-gluon glueballs are expected to couple predominantly to light hadrons, with suppressed decay rates into heavy quarkonium.

In addition, the production mechanisms of these two types of states are expected to be markedly different. Heavy tetraquark hybrid states are favored in processes associated with heavy-flavor production, such as heavy quarkonium decays and high-energy hadronic collisions, whereas three-gluon glueballs are more naturally produced in gluon-rich environments without heavy-flavor enhancement. These differences in mass scales, internal structures, and production and decay properties provide viable experimental criteria for disentangling tetraquark hybrid states from possible glueball candidates.

Finally, the tetraquark hybrid states in the bottom sector are predicted to be significantly heavier than their charm counterparts, rendering their experimental production more challenging. Moreover, one of the proposed production mechanisms proceeds through $\Upsilon$ decays, which is not applicable to the hidden-bottom tetraquark hybrid states themselves. As a consequence, experimental searches for hidden-bottom tetraquark hybrid states are expected to be considerably more difficult than those for the hidden-charm sector.

\vspace{0.7cm} {\bf Acknowledgments}

This work was supported in part by the National Natural Science Foundation of China under Grants 12575106 and 12147214, and Specific Fund of Fundamental Scientific Research Operating Expenses for Undergraduate Universities in Liaoning Province under Grants No. LJ212410165019.

\begin{widetext}
\appendix
\section{The spectral densities for gluonic tetraquark states}\label{App_A}
\subsection{The spectral densities for $0^{++}$ gluonic tetraquark states}
For the current shown in Eq. (\ref{current-0++A}),  we obtain the spectral densities as follows:
\begin{eqnarray}
\rho^{pert}_{0^{++}\;,A} (s) &=& \int^{\alpha_{max}}_{\alpha_{min}} d \alpha \int^{1 - \alpha}_{\beta_{min}} d \beta \frac{g_s^2{\cal F}^5_{\alpha \beta} (\alpha + \beta - 1)^3 \big({\cal F}_{\alpha\beta}+3m_c m_q(\alpha +\beta )\big)}{5\times 3^2 \times 2^{11}\pi^8 \alpha^5 \beta^5}\; , \\
\rho_{0^{++}\;,A}^{\langle \bar{q} q \rangle}(s) &=& \frac{\langle \bar{q} q \rangle}{\pi^6}
\int^{\alpha_{max}}_{\alpha_{min}} d \alpha \int^{1 - \alpha}_{\beta_{min}} d \beta
\frac{ g_s^2 {\cal F}_{\alpha \beta }^3(\alpha+\beta-1)}{3 \times 2^9 \alpha^4 \beta^4} \big( 4 \alpha\beta{\cal F}_{\alpha \beta} m_q\nonumber\\
&+&4\alpha\beta m_c m_q^2(\alpha+\beta)-m_c{\cal F}_{\alpha\beta}(\alpha+\beta)(\alpha+\beta-1)\big)\; , \\
\rho_{0^{++}\;,A}^{\langle G^2 \rangle\;,I}(s) &=& \frac{\langle g_s^2 G^2\rangle}{\pi^6} \int^{\alpha_{max}}_{\alpha_{min}} d \alpha \int^{1 - \alpha}_{\beta_{min}} d \beta \frac{m_c{\cal F}_{\alpha \beta}^2(\alpha+\beta-1)}{3 \times 2^{9} \alpha^3 \beta^3} \big({\cal F}_{\alpha\beta}m_q(\alpha \nonumber\\
&+&\beta )+6\alpha\beta m_c m_q^2\big)
 \; , \\
 \rho_{0^{++}\;,A}^{\langle G^2 \rangle\;,II}(s) &=& \frac{g_s^2 \langle g_s^2 G^2\rangle}{\pi^8} \int^{\alpha_{max}}_{\alpha_{min}} d \alpha \int^{1 - \alpha}_{\beta_{min}} d \beta \bigg( \frac{m_c{\cal F}_{\alpha \beta}^2(\alpha+\beta-1)^3}{3^3 \times 2^{12} \alpha^5 \beta^5} \big( \alpha^3{\cal F}_{\alpha\beta}(2 m_c+3 m_q)\nonumber\\
&+&\beta^3 {\cal F}_{\alpha\beta}(2 m_c+3 m_q) + 3m_q m_c^2(\alpha^3+\beta^3)(\alpha + \beta )\big)+\frac{{\cal F}_{\alpha \beta}^3(\alpha+\beta-1)}{3^3 \times 2^{19} \alpha^4 \beta^4} \nonumber\\
&\times&\big( {\cal F}_{\alpha\beta}(1-5\alpha+4\alpha^2-5\beta+4\beta^2+14\alpha\beta)+ 4m_c(\alpha+\beta-1)(\beta m_q(49\alpha+4\beta\nonumber\\
&-&1) +\alpha m_q(49\beta+4\alpha-1)) \big)\bigg)
 \; , \\
\rho_{0^{++}\;,A}^{\langle \bar{q} G q \rangle}(s) &=& \frac{\langle g_s \bar{q} \sigma \cdot G q \rangle}{\pi^6} \int_{\alpha_{min}}^{\alpha_{max}}  \int_{\beta_{min}}^{1 - \alpha} d \beta  \frac{g_s^2 {\cal F}^2_{\alpha\beta}}{3^2 \times 2^{13} \alpha^3 \beta^3}\big( -128\alpha\beta {\cal F}_{\alpha\beta}m_q \nonumber\\
&-&96\alpha\beta(\alpha+\beta)m_c m_q^2 + m_c {\cal F}_{\alpha\beta} (\alpha+\beta-1)(94\alpha+94\beta+1)\big)\; ,\\
\rho_{0^{++}\;,A}^{\langle \bar{q} q\rangle^2}(s) &=& \frac{\langle \bar{q} q\rangle^2}{\pi^4} \int_{\alpha_{min}}^{\alpha_{max}} d \alpha \bigg{\{}\int_{\beta_{min}}^{1 - \alpha} d \beta \big[ -\frac{g_s^2 m_c m_q {\cal F}_{\alpha\beta}^2(\alpha +\beta ) }{3 \times 2^{5} \alpha^2\beta^2}  \big] \nonumber\\
&+&\big[\frac{g_s^2 {\cal H}_\alpha^2m_q^2}{3\times 2^5\alpha(\alpha-1)}\big]\bigg{\}}\; , \\
\rho_{0^{++}\;,A}^{\langle G^3 \rangle\;,I}(s) &=& \frac{\langle g_s^3 G^3 \rangle}{\pi^6} \int_{\alpha_{min}}^{\alpha_{max}} d\alpha \int_{\beta_{min}}^{1 - \alpha} d \beta - \frac{{\cal F}_{\alpha\beta}}{3 \times 2^{11} \alpha^3 \beta^3} \big[ 2 \alpha\beta {\cal F}_{\alpha\beta}^2 -12\alpha\beta m_q^2 m_c^2 (\alpha^2-\beta^2\nonumber\\
&-&\alpha\beta+\beta-\alpha)  - 3m_c{\cal F}_{\alpha\beta} \big(\alpha m_q(\alpha^2-\beta^2
-3\alpha\beta+\beta-\alpha)+\beta m_q(\alpha^2-\beta^2\nonumber\\
&-&\alpha\beta+\beta-\alpha)\big)   \big] \; , \\
\rho_{0^{++}\;,A}^{\langle G^3 \rangle\;,II}(s) &=& \frac{\langle g_s^3 G^3 \rangle}{\pi^8} \int_{\alpha_{min}}^{\alpha_{max}} d\alpha \int_{\beta_{min}}^{1 - \alpha} d \beta  \frac{g_s^2 {\cal F}_{\alpha\beta}(\alpha+\beta-1)^3}{3^3 \times 2^{14} \alpha^5 \beta^5} \big[2{\cal F}_{\alpha\beta}^2(\alpha^3+\beta^3)\nonumber\\
&+&12m_c^3m_q(\alpha +\beta )(\alpha^4+\beta^4)+3m_c {\cal F}_{\alpha\beta}\big(6\alpha^4 m_q+6\beta^4 m_q \nonumber\\
&+&\alpha\beta m_q(\alpha^2  +\beta^2 )+4m_c(\alpha^4+\beta^4)\big)  \big] \; , \\
\rho_{0^{++}\;,A}^{\langle \bar{q} q\rangle\langle \bar{q} G q \rangle}(s) &=& \frac{g_s^2\langle \bar{q} q\rangle \langle g_s \bar{q} \sigma \cdot G q \rangle}{\pi^4} \int_{\alpha_{min}}^{\alpha_{max}} d \alpha \bigg{\{}\int_{\beta_{min}}^{1 - \alpha} d \beta \big[ -\frac{ m_c m_q {\cal F}_{\alpha\beta}}{3^2 \times 2^{11} \alpha\beta}  \big] \nonumber\\
&+&\frac{1}{3^2\times 2^6\alpha(\alpha-1)}\big[-12{\cal H}_\alpha \alpha(\alpha-1) m_q^2 +4\alpha (\alpha-1)m_c^2 m_q^2 \nonumber\\
&+&5m_c {\cal H}_\alpha m_q  \big]\bigg{\}}\; , 
\end{eqnarray}
where, the subscript $I$ and $II$ for the gluon condensate come from dynamic gluon and quarks, respectively. Here,
\begin{eqnarray}
{\cal F}_{\alpha \beta} &=& (\alpha + \beta) m_c^2 - \alpha \beta s \; , {\cal H}_\alpha  = m_c^2 - \alpha (1 - \alpha) s \; , \\
\alpha_{min} &=& \left(1 - \sqrt{1 - 4 m_c^2/s} \right) / 2, \; , \alpha_{max} = \left(1 + \sqrt{1 - 4 m_c^2 / s} \right) / 2  \; , \\
\beta_{min} &=& \alpha m_c^2 /(s \alpha - m_c^2).
\end{eqnarray}

For the current shown in Eq. (\ref{current-0++B}), we obtain the spectral densities as follows:
\begin{eqnarray}
\rho^{pert}_{0^{++}\;,B} (s) &=& \int^{\alpha_{max}}_{\alpha_{min}} d \alpha \int^{1 - \alpha}_{\beta_{min}} d \beta \frac{g_s^2{\cal F}^5_{\alpha \beta} (\alpha + \beta - 1)^3 \big({\cal F}_{\alpha\beta}-3m_c m_q (\alpha +\beta )\big)}{5\times 3^2 \times 2^{11}\pi^8 \alpha^5 \beta^5}\; ,  \\
\rho_{0^{++}\;,B}^{\langle \bar{q} q \rangle}(s) &=& \frac{\langle \bar{q} q \rangle}{\pi^6}
\int^{\alpha_{max}}_{\alpha_{min}} d \alpha \int^{1 - \alpha}_{\beta_{min}} d \beta
\frac{ g_s^2 {\cal F}_{\alpha \beta }^3(\alpha+\beta-1)}{3 \times 2^9 \alpha^4 \beta^4} \big( 4 \alpha\beta{\cal F}_{\alpha \beta} m_q\nonumber\\
&-&4\alpha\beta m_c m_q^2(\alpha+\beta)+m_c{\cal F}_{\alpha\beta}(\alpha+\beta)(\alpha+\beta-1)\big)\; , \\
\rho_{0^{++}\;,B}^{\langle G^2 \rangle\;,I}(s) &=& \frac{\langle g_s^2 G^2\rangle}{\pi^6} \int^{\alpha_{max}}_{\alpha_{min}} d \alpha \int^{1 - \alpha}_{\beta_{min}} d \beta - \frac{m_c{\cal F}_{\alpha \beta}^2(\alpha+\beta-1)}{3 \times 2^{9} \alpha^3 \beta^3} \big({\cal F}_{\alpha\beta}m_q(\alpha \nonumber\\
&+&\beta )-6\alpha\beta m_c m_q^2\big)
 \; , \\
 \rho_{0^{++}\;,B}^{\langle G^2 \rangle\;,II}(s) &=& \frac{g_s^2 \langle g_s^2 G^2\rangle}{\pi^8} \int^{\alpha_{max}}_{\alpha_{min}} d \alpha \int^{1 - \alpha}_{\beta_{min}} d \beta \bigg(- \frac{m_c{\cal F}_{\alpha \beta}^2(\alpha+\beta-1)^3}{3^3 \times 2^{12} \alpha^5 \beta^5} \big( \alpha^3{\cal F}_{\alpha\beta}(3 m_q\nonumber\\
 &-&2 m_c)+\beta^3 {\cal F}_{\alpha\beta}(3 m_q-2 m_c) + 3m_c^2 m_q (\alpha^3+\beta^3)(\alpha + \beta )\big)\nonumber\\
&+&\frac{{\cal F}_{\alpha \beta}^3(\alpha+\beta-1)}{3^3 \times 2^{19} \alpha^4 \beta^4} \big( -{\cal F}_{\alpha\beta}(1-5\alpha+4\alpha^2-5\beta+4\beta^2+14\alpha\beta)\nonumber\\
&+& 4m_c(\alpha+\beta-1)(\beta m_q(49\alpha+4\beta-1) +\alpha m_q(49\beta+4\alpha-1)) \big)\bigg)
 \; , \\
\rho_{0^{++}\;,B}^{\langle \bar{q} G q \rangle}(s) &=& -\frac{\langle g_s \bar{q} \sigma \cdot G q \rangle}{\pi^6} \int_{\alpha_{min}}^{\alpha_{max}}  \int_{\beta_{min}}^{1 - \alpha} d \beta  \frac{g_s^2 {\cal F}^2_{\alpha\beta}}{3^2 \times 2^{13} \alpha^3 \beta^3}\big( 128 m_q \alpha\beta {\cal F}_{\alpha\beta} \nonumber\\
&-&96\alpha\beta(\alpha+\beta)m_c m_q^2 + m_c {\cal F}_{\alpha\beta}(\alpha+\beta-1)(94\alpha+94\beta+1) \big)\; ,\\
\rho_{0^{++}\;,B}^{\langle \bar{q} q\rangle^2}(s) &=& \frac{\langle \bar{q} q\rangle^2}{\pi^4} \int_{\alpha_{min}}^{\alpha_{max}} d \alpha \bigg{\{}\int_{\beta_{min}}^{1 - \alpha} d \beta \big[ \frac{g_s^2 m_c {\cal F}_{\alpha\beta}^2 m_q (\alpha +\beta) }{3 \times 2^{5} \alpha^2\beta^2}  \big] \nonumber\\
&+&\big[\frac{g_s^2 {\cal H}_\alpha^2m_q^2}{3\times 2^5\alpha(\alpha-1)}\big]\bigg{\}}\; , \\
\rho_{0^{++}\;,B}^{\langle G^3 \rangle\;,I}(s) &=& \frac{\langle g_s^3 G^3 \rangle}{\pi^6} \int_{\alpha_{min}}^{\alpha_{max}} d\alpha \int_{\beta_{min}}^{1 - \alpha} d \beta - \frac{{\cal F}_{\alpha\beta}}{3 \times 2^{11} \alpha^3 \beta^3} \big[ 2 \alpha\beta {\cal F}_{\alpha\beta}^2 \nonumber\\
&-&12\alpha\beta m_q^2 m_c^2 (\alpha^2-\beta^2-\alpha\beta+\beta-\alpha)  + 3m_c{\cal F}_{\alpha\beta} \big(\alpha m_q(\alpha^2-\beta^2\nonumber\\
&-&3\alpha\beta+\beta-\alpha)+\beta m_q(\alpha^2-\beta^2-\alpha\beta+\beta-\alpha)\big)   \big] \; , \\
\rho_{0^{++}\;,B}^{\langle G^3 \rangle\;,II}(s) &=& \frac{\langle g_s^3 G^3 \rangle}{\pi^8} \int_{\alpha_{min}}^{\alpha_{max}} d\alpha \int_{\beta_{min}}^{1 - \alpha} d \beta - \frac{g_s^2 {\cal F}_{\alpha\beta}(\alpha+\beta-1)^3}{3^3 \times 2^{14} \alpha^5 \beta^5} \big[-2{\cal F}_{\alpha\beta}^2(\alpha^3+\beta^3)\nonumber\\
&+&12m_c^3 m_q (\alpha +\beta )(\alpha^4+\beta^4)+3m_c {\cal F}_{\alpha\beta}\big(6\alpha^4 m_q+6\beta^4 m_q \nonumber\\
&+&m_q\alpha\beta(\alpha^2  +\beta^2 )-4m_c(\alpha^4+\beta^4)\big)  \big] \; , \\
\rho_{0^+\;,B}^{\langle \bar{q} q\rangle\langle \bar{q} G q \rangle}(s) &=& \frac{g_s^2\langle \bar{q} q\rangle \langle g_s \bar{q} \sigma \cdot G q \rangle}{\pi^4} \int_{\alpha_{min}}^{\alpha_{max}} d \alpha \bigg{\{}\int_{\beta_{min}}^{1 - \alpha} d \beta \big[ \frac{ m_c {\cal F}_{\alpha\beta} m_q) }{3^2 \times 2^{10} \alpha\beta}  \big] \nonumber\\
&+&\frac{1}{3^2\times 2^6\alpha(\alpha-1)}\big[-12{\cal H}_\alpha \alpha(\alpha-1) m_q^2 +4\alpha (\alpha-1)m_c^2 m_q^2 \nonumber\\
&-&5m_c {\cal H}_\alpha m_q  \big]\bigg{\}}\; .
\end{eqnarray}

For the current showned in Eq. (\ref{current-0++C}), we obtain the spectral densities as follows:
\begin{eqnarray}
\rho^{pert}_{0^{++}\;,C} (s) &=& \int^{\alpha_{max}}_{\alpha_{min}} d \alpha \int^{1 - \alpha}_{\beta_{min}} d \beta \frac{g_s^2{\cal F}^6_{\alpha \beta} (\alpha + \beta - 1)^3 }{5\times 3^2 \times 2^{11}\pi^8 \alpha^5 \beta^5}\; , \\
\rho_{0^{++}\;,C}^{\langle \bar{q} q \rangle}(s) &=& \frac{\langle \bar{q} q \rangle}{\pi^6}
\int^{\alpha_{max}}_{\alpha_{min}} d \alpha \int^{1 - \alpha}_{\beta_{min}} d \beta
\frac{ g_s^2 m_q {\cal F}_{\alpha \beta }^4(\alpha+\beta-1)}{3 \times 2^{7} \alpha^3 \beta^3}\; , \\
\rho_{0^{++}\;,C}^{\langle G^2 \rangle\;,I}(s) &=& \frac{\langle g_s^2 G^2\rangle}{\pi^6} \int^{\alpha_{max}}_{\alpha_{min}} d \alpha \int^{1 - \alpha}_{\beta_{min}} d \beta \frac{m_c^2 m_q^2{\cal F}_{\alpha \beta}^2(\alpha+\beta-1)}{  2^{8} \alpha^2 \beta^2}  \; , \\
 \rho_{0^{++}\;,C}^{\langle G^2 \rangle\;,II}(s) &=& \frac{g_s^2 \langle g_s^2 G^2\rangle}{\pi^8} \int^{\alpha_{max}}_{\alpha_{min}} d \alpha \int^{1 - \alpha}_{\beta_{min}} d \beta \bigg( \frac{m_c^2{\cal F}_{\alpha \beta}^3(\alpha+\beta-1)^3(\alpha^3+\beta^3)}{3^3 \times 2^{11} \alpha^5 \beta^5} \nonumber\\
&+&\frac{{\cal F}_{\alpha \beta}^4(\alpha+\beta-1)(\alpha+\beta)}{3^2 \times 2^{18} \alpha^4 \beta^4} \bigg)\; , \\
\rho_{0^{++}\;,C}^{\langle \bar{q} G q \rangle}(s) &=&- \frac{\langle g_s \bar{q} \sigma \cdot G q \rangle}{\pi^6} \int_{\alpha_{min}}^{\alpha_{max}}  \int_{\beta_{min}}^{1 - \alpha} d \beta  \frac{g_s^2 m_q {\cal F}^3_{\alpha\beta}}{3^2 \times 2^{6} \alpha^2 \beta^2}\; ,\\
\rho_{0^{++}\;,C}^{\langle \bar{q} q\rangle^2}(s) &=& \frac{\langle \bar{q} q\rangle^2}{\pi^4} \int_{\alpha_{min}}^{\alpha_{max}} d \alpha \bigg{\{}\frac{g_s^2 {\cal H}_\alpha^2m_q^2}{3\times 2^5\alpha(\alpha-1)}\bigg{\}}\; , \\
\rho_{0^{++}\;,C}^{\langle G^3 \rangle\;,I}(s) &=& \frac{\langle g_s^3 G^3 \rangle}{\pi^6} \int_{\alpha_{min}}^{\alpha_{max}} d\alpha \int_{\beta_{min}}^{1 - \alpha} d \beta - \frac{1}{3 \times 2^{10} \alpha^2 \beta^2} \big[  {\cal F}_{\alpha\beta}^3 \nonumber\\
&+&6 m_q^2 m_c^2 {\cal F}_{\alpha\beta}(-4\alpha^2+4\beta^2+\alpha\beta-4\beta+4\alpha)    \big] \; , \\
\rho_{0^{++}\;,C}^{\langle G^3 \rangle\;,II}(s) &=& \frac{\langle g_s^3 G^3 \rangle}{\pi^8} \int_{\alpha_{min}}^{\alpha_{max}} d\alpha \int_{\beta_{min}}^{1 - \alpha} d \beta  \frac{g_s^2 {\cal F}_{\alpha\beta}^2(\alpha+\beta-1)^3}{3^3 \times 2^{13} \alpha^5 \beta^5} \big[{\cal F}_{\alpha\beta}(\alpha^3+\beta^3)\nonumber\\
&+&6 m_c^2 (\alpha^4+\beta^4)  \big] \; , \\
\rho_{0^{++}\;,C}^{\langle \bar{q} q\rangle\langle \bar{q} G q \rangle}(s) &=& \frac{g_s^2\langle \bar{q} q\rangle \langle g_s \bar{q} \sigma \cdot G q \rangle}{\pi^4} \int_{\alpha_{min}}^{\alpha_{max}} d \alpha \bigg{\{}
\frac{(m_c^2-3{\cal H}_\alpha)m_q^2}{3^2\times 2^4}\bigg{\}}\; .
\end{eqnarray}

\subsection{The spectral densities for $0^{-+}$ gluonic tetraquark states}

For the current shown in Eq. (\ref{current-0-+A}),  we obtain the spectral densities as follows:
\begin{eqnarray}
\rho^{pert}_{0^{-+}\;,A} (s) &=& \rho^{pert}_{0^{++}\;,A} (s)\; , \\
\rho_{0^{-+}\;,A}^{\langle \bar{q} q \rangle}(s) &=& \rho_{0^{++}\;,A}^{\langle \bar{q} q \rangle}(s)\; , \\
\rho_{0^{-+}\;,A}^{\langle G^2 \rangle\;,I}(s) &=&-\rho_{0^{++}\;,A}^{\langle G^2 \rangle\;,I}(s)
 \; , \\
 \rho_{0^{-+}\;,A}^{\langle G^2 \rangle\;,II}(s) &=& \rho_{0^{++}\;,A}^{\langle G^2 \rangle\;,II}(s)
 \; , \\
\rho_{0^{-+}\;,A}^{\langle \bar{q} G q \rangle}(s) &=&\rho_{0^{++}\;,A}^{\langle \bar{q} G q \rangle}(s)\; ,\\
\rho_{0^{-+}\;,A}^{\langle \bar{q} q\rangle^2}(s) &=& \rho_{0^{++}\;,A}^{\langle \bar{q} q\rangle^2}(s)\; , \\
\rho_{0^{-+}\;,A}^{\langle G^3 \rangle\;,I}(s) &=& \frac{\langle g_s^3 G^3 \rangle}{\pi^6} \int_{\alpha_{min}}^{\alpha_{max}} d\alpha \int_{\beta_{min}}^{1 - \alpha} d \beta - \frac{{\cal F}_{\alpha\beta}}{3 \times 2^{10} \alpha^3 \beta^3} \big[  \alpha\beta {\cal F}_{\alpha\beta}^2 \nonumber\\
&+&6\alpha\beta m_q^2 m_c^2 (4\alpha^2-4\beta^2-\alpha\beta+\beta-\alpha)  + 6m_c{\cal F}_{\alpha\beta} \big(\alpha m_q(\alpha^2\nonumber\\
&-&\beta^2-\alpha\beta+\beta-\alpha)+\beta m_q(\alpha^2-\beta^2+\alpha\beta+\beta-\alpha)\big)   \big] \; , \\
\rho_{0^{-+}\;,A}^{\langle G^3 \rangle\;,II}(s) &=& \rho_{0^{++}\;,A}^{\langle G^3 \rangle\;,II}(s) \; , \\
\rho_{0^{-+}\;,A}^{\langle \bar{q} q\rangle\langle \bar{q} G q \rangle}(s) &=&\rho_{0^{++}\;,A}^{\langle \bar{q} q\rangle\langle \bar{q} G q \rangle}(s)\; .
\end{eqnarray}

For the currents shown in Eq. (\ref{current-0-+B}), we obtain the spectral densities as follows:
\begin{eqnarray}
\rho^{pert}_{0^{-+}\;,B} (s) &=& \rho^{pert}_{0^{++}\;,B} (s) \; , \\
\rho_{0^{-+}\;,B}^{\langle \bar{q} q \rangle}(s) &=& \rho_{0^{++}\;,B}^{\langle \bar{q} q \rangle}(s) \; , \\
\rho_{0^{-+}\;,B}^{\langle G^2 \rangle\;,I}(s) &=& -\rho_{0^{++}\;,B}^{\langle G^2 \rangle\;,I}(s)
 \; , \\
 \rho_{0^{-+}\;,B}^{\langle G^2 \rangle\;,II}(s) &=&  \rho_{0^{++}\;,B}^{\langle G^2 \rangle\;,II}(s)
 \; , \\
\rho_{0^{-+}\;,B}^{\langle \bar{q} G q \rangle}(s) &=& \rho_{0^{++}\;,B}^{\langle \bar{q} G q \rangle}(s) \; ,\\
\rho_{0^{-+}\;,B}^{\langle \bar{q} q\rangle^2}(s) &=& \rho_{0^{++}\;,B}^{\langle \bar{q} q\rangle^2}(s) \; , \\
\rho_{0^{-+}\;,B}^{\langle G^3 \rangle\;,I}(s) &=& \frac{\langle g_s^3 G^3 \rangle}{\pi^6} \int_{\alpha_{min}}^{\alpha_{max}} d\alpha \int_{\beta_{min}}^{1 - \alpha} d \beta - \frac{{\cal F}_{\alpha\beta}}{3 \times 2^{10} \alpha^3 \beta^3} \big[  \alpha\beta {\cal F}_{\alpha\beta}^2 +6\alpha\beta m_q^2 m_c^2 (4\alpha^2-4\beta^2\nonumber\\
&-&\alpha\beta+4\beta-4\alpha)  - 6m_c{\cal F}_{\alpha\beta} \big(\alpha m_q(\alpha^2-\beta^2
-\alpha\beta+\beta-\alpha)+\beta m_q(\alpha^2-\beta^2\nonumber\\
&+&\alpha\beta+\beta-\alpha)\big)   \big] \; , \\
\rho_{0^{-+}\;,B}^{\langle G^3 \rangle\;,II}(s) &=& \rho_{0^{++}\;,B}^{\langle G^3 \rangle\;,II}(s)  \; , \\
\rho_{0^{-+}\;,B}^{\langle \bar{q} q\rangle\langle \bar{q} G q \rangle}(s) &=& \rho_{0^{++}\;,B}^{\langle \bar{q} q\rangle\langle \bar{q} G q \rangle}(s)\; .
\end{eqnarray}

For the current showned in Eq. (\ref{current-0-+C}), we obtain the spectral densities as follows:
\begin{eqnarray}
\rho^{pert}_{0^{-+}\;,C} (s) &=&\rho^{pert}_{0^{++}\;,C} (s) \; ,  \\
\rho_{0^{-+}\;,C}^{\langle \bar{q} q \rangle}(s) &=& \rho_{0^{++}\;,C}^{\langle \bar{q} q \rangle}(s)\; , \\
\rho_{0^{-+}\;,C}^{\langle G^2 \rangle\;,I}(s) &=&-\rho_{0^{++}\;,C}^{\langle G^2 \rangle\;,I}(s)
 \; , \\
 \rho_{0^{-+}\;,C}^{\langle G^2 \rangle\;,II}(s) &=& \rho_{0^{++}\;,C}^{\langle G^2 \rangle\;,II}(s) 
 \; , \\
\rho_{0^{-+}\;,C}^{\langle \bar{q} G q \rangle}(s) &=&\rho_{0^{++}\;,C}^{\langle \bar{q} G q \rangle}(s)\; ,\\
\rho_{0^{-+}\;,C}^{\langle \bar{q} q\rangle^2}(s) &=& \rho_{0^{++}\;,C}^{\langle \bar{q} q\rangle^2}(s)\; , \\
\rho_{0^{-+}\;,C}^{\langle G^3 \rangle\;,I}(s) &=&\frac{\langle g_s^3 G^3 \rangle}{\pi^6} \int_{\alpha_{min}}^{\alpha_{max}} d\alpha \int_{\beta_{min}}^{1 - \alpha} d \beta - \frac{1}{3 \times 2^{10} \alpha^2 \beta^2} \big[  {\cal F}_{\alpha\beta}^3 \nonumber\\
&+&6 m_q^2 m_c^2 {\cal F}_{\alpha\beta}(\alpha^2-\beta^2-\alpha\beta+\beta-\alpha)    \big] \; , \\
\rho_{0^{-+}\;,C}^{\langle G^3 \rangle\;,II}(s) &=& \rho_{0^{++}\;,C}^{\langle G^3 \rangle\;,II}(s)  \; , \\
\rho_{0^{-+}\;,C}^{\langle \bar{q} q\rangle\langle \bar{q} G q \rangle}(s) &=& \rho_{0^{++}\;,C}^{\langle \bar{q} q\rangle\langle \bar{q} G q \rangle}(s)\; .
\end{eqnarray}

\subsection{The spectral densities for $0^{--}$ gluonic tetraquark states}

For the current shown in Eq. (\ref{current-0--A}),  we obtain the spectral densities as follows:
\begin{eqnarray}
\rho^{pert}_{0^{--}\;,A} (s) &=& \rho^{pert}_{0^{-+}\;,C} (s) \; ,  \\
\rho_{0^{--}\;,A}^{\langle \bar{q} q \rangle}(s) &=& \rho_{0^{-+}\;,C}^{\langle \bar{q} q \rangle}(s)\; , \\
\rho_{0^{--}\;,A}^{\langle G^2 \rangle\;,I}(s) &=& \rho_{0^{-+}\;,C}^{\langle G^2 \rangle\;,I}(s) \; , \\
 \rho_{0^{--}\;,A}^{\langle G^2 \rangle\;,II}(s) &=& \frac{g_s^2 \langle g_s^2 G^2\rangle}{\pi^8} \int^{\alpha_{max}}_{\alpha_{min}} d \alpha \int^{1 - \alpha}_{\beta_{min}} d \beta \bigg( \frac{m_c^2{\cal F}_{\alpha \beta}^3(\alpha+\beta-1)^3(\alpha^3+\beta^3)}{3^3 \times 2^{11} \alpha^5 \beta^5} \nonumber\\
&+&\frac{{\cal F}_{\alpha \beta}^4(\alpha+\beta-1)(\alpha^2+(\beta-1)^2-2\alpha+8\alpha\beta)}{3^3 \times 2^{18} \alpha^4 \beta^4} \bigg) \; , \\
\rho_{0^{--}\;,A}^{\langle \bar{q} G q \rangle}(s) &=& \rho_{0^{-+}\;,C}^{\langle \bar{q} G q \rangle}(s)\; ,\\
\rho_{0^{--}\;,A}^{\langle \bar{q} q\rangle^2}(s) &=&\rho_{0^{-+}\;,C}^{\langle \bar{q} q\rangle^2}(s) \; , \\
\rho_{0^{--}\;,A}^{\langle G^3 \rangle\;,I}(s) &=& \rho_{0^{-+}\;,C}^{\langle G^3 \rangle\;,I}(s) \; , \\
\rho_{0^{--}\;,A}^{\langle G^3 \rangle\;,II}(s) &=&\rho_{0^{-+}\;,C}^{\langle G^3 \rangle\;,II}(s)\; , \\
\rho_{0^{--}\;,A}^{\langle \bar{q} q\rangle\langle \bar{q} G q \rangle}(s) &=& \rho_{0^{-+}\;,C}^{\langle \bar{q} q\rangle\langle \bar{q} G q \rangle}(s) \; .
\end{eqnarray}

For the currents shown in Eq. (\ref{current-0+-A}), we obtain the spectral densities as follows:
\begin{eqnarray}
\rho^{pert}_{0^{+-}\;,A} (s) &=& \rho^{pert}_{0^{--}\;,A} (s)\; ,  \\
\rho_{0^{+-}\;,A}^{\langle \bar{q} q \rangle}(s) &=& \rho_{0^{--}\;,A}^{\langle \bar{q} q \rangle}(s)\; , \\
\rho_{0^{+-}\;,A}^{\langle G^2 \rangle\;,I}(s) &=& -\rho_{0^{--}\;,A}^{\langle G^2 \rangle\;,I}(s)\; , \\
\rho_{0^{+-}\;,A}^{\langle G^2 \rangle\;,II}(s) &=&\rho_{0^{--}\;,A}^{\langle G^2 \rangle\;,II}(s) \; , \\
\rho_{0^{+-}\;,A}^{\langle \bar{q} G q \rangle}(s) &=& \rho_{0^{--}\;,A}^{\langle \bar{q} G q \rangle}(s)\; ,\\
\rho_{0^{+-}\;,A}^{\langle \bar{q} q\rangle^2}(s) &=&\rho_{0^{--}\;,A}^{\langle \bar{q} q\rangle^2}(s) \; , \\
\rho_{0^{+-}\;,B}^{\langle G^3 \rangle\;,I}(s) &=& \frac{\langle g_s^3 G^3 \rangle}{\pi^6} \int_{\alpha_{min}}^{\alpha_{max}} d\alpha \int_{\beta_{min}}^{1 - \alpha} d \beta - \frac{1}{3 \times 2^{10} \alpha^2 \beta^2} \big[  {\cal F}_{\alpha\beta}^3 \nonumber\\
&+&6 m_q^2 m_c^2 {\cal F}_{\alpha\beta}(-4\alpha^2+4\beta^2+\alpha\beta-\beta+4\alpha)    \big]\; , \\
\rho_{0^{+-}\;,A}^{\langle G^3 \rangle\;,II}(s) &=& \rho_{0^{--}\;,A}^{\langle G^3 \rangle\;,II}(s) \; , \\
\rho_{0^{+-}\;,A}^{\langle \bar{q} q\rangle\langle \bar{q} G q \rangle}(s) &=&\rho_{0^{--}\;,A}^{\langle \bar{q} q\rangle\langle \bar{q} G q \rangle}(s)\; .
\end{eqnarray}

\section{An Example of Wick Contractions for the Gluonic Current}

In this appendix, we present a representative example of the Wick contractions for the gluonic interpolating current used in this work.

The current is given by
\begin{align}
j^{A}_{0^{++}} (x) &= g_s \epsilon_{ikl}\epsilon_{jmn} 
\left[q_k^T(x) C \gamma_\mu c_l(x)\right]
\frac{\lambda_{ij}^a}{2} G_{\mu\nu}^a(x)
\left[\bar{q}_m(x) \gamma_\nu C \bar{c}_n^{T}(x)\right].
\end{align}

The two-point correlation function reads
\begin{align}
\Pi(q) = i \int d^4x \, e^{iqx} 
\langle 0 | T[j^{A}(x) j^{A\dagger}(0)] | 0 \rangle.
\end{align}

Applying Wick's theorem, the quark fields are contracted into propagators:
\begin{align}
\langle 0 | T \, q_k^T(x) \bar{q}_{m}(0) | 0 \rangle 
&= C \, S_q^{T}(x)_{km}, \\
\langle 0 | T \, c_l(x) \bar{c}_n(0) | 0 \rangle 
&= S_c(x)_{ln}.
\end{align}

As a result, the correlation function can be schematically written as
\begin{align}
\Pi(q) \sim g_s^2 \int d^4x \, e^{iqx} \,
\epsilon_{ikl}\epsilon_{jmn} \epsilon_{i'k'l'}\epsilon_{j'm'n'} 
\frac{\lambda_{ij}^a}{2} \frac{\lambda_{i'j'}^{a'}}{2}
\langle G_{\mu\nu}^a(x) G_{\rho\sigma}^{a'}(0) \rangle \nonumber\\
\times \text{Tr}\left[\gamma_\mu S_c(x) \gamma_\rho S_q(-x)\right]
\text{Tr}\left[\gamma_\nu S_c(-x) \gamma_\sigma S_q(x)\right] + \cdots.
\end{align}

The gluonic fields are treated within the operator product expansion. At leading order, their contraction leads to the gluon condensate contribution:
\begin{align}
\langle G_{\mu\nu}^a(x) G_{\rho\sigma}^{a'}(0) \rangle 
\sim \delta^{aa'} \langle G^2 \rangle + \cdots,
\end{align}
while higher-order terms generate contributions from higher-dimensional condensates.

This example illustrates how the interpolating current with explicit gluonic degrees of freedom is systematically reduced to quark propagators and gluon condensates in the QCD sum rule framework.

\section{The ratios $R^{OPE}$, $R^{PC}$, and the masses $m$ are plopted as functions of Borel parameter $M_B^2$}\label{App_B}

We display the figures of the $R^{OPE}$, $R^{PC}$, and the masses $m$ as functions of Borel parameter $M_B^2$ for hidden-charm and -bottom tetraquark hybrid states below. It should be noted that the differences between the numerical results for $j_{0^+}^C$ and $j_{0^+}^D$ are so tiny that we just plot the case of $j_{0^+}^C$. The same applies to the pairs $(j_{0^-}^C, j_{0^-}^D)$, $(j_{1^-}^B, j_{1^-}^F)$, $(j_{1^-}^C, j_{1^-}^G)$, and $(j_{1^+}^C, j_{1^+}^G)$. 

\begin{figure}
\includegraphics[width=6.8cm]{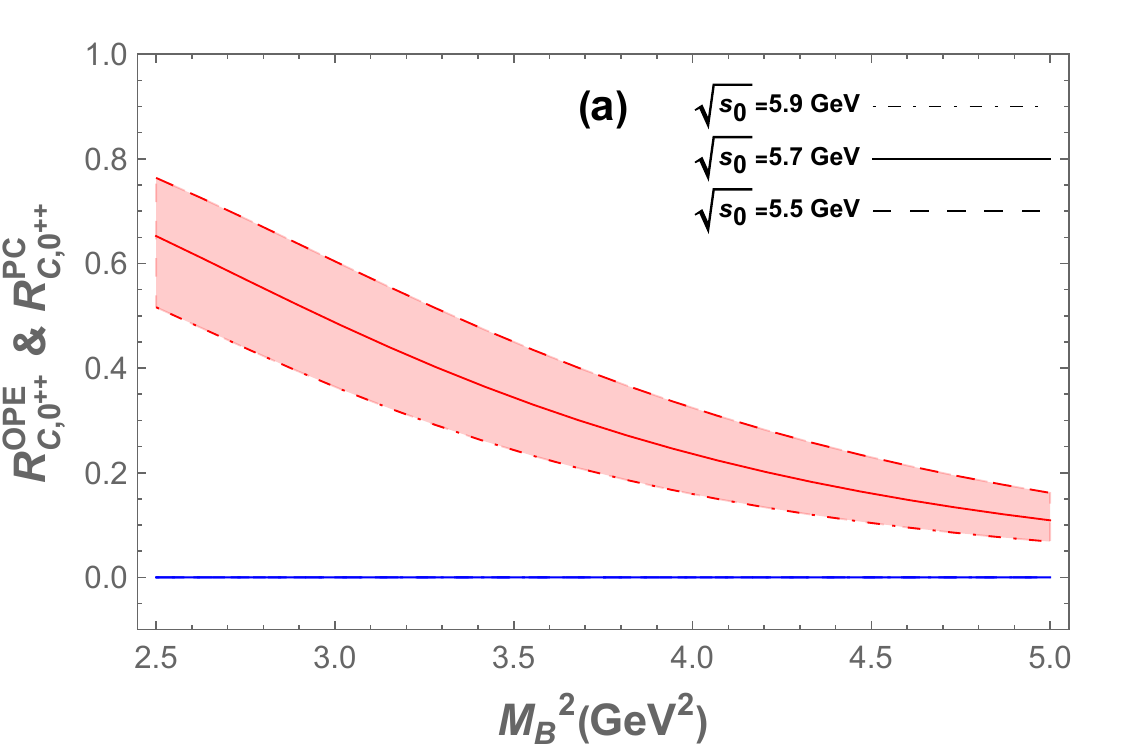}
\includegraphics[width=6.8cm]{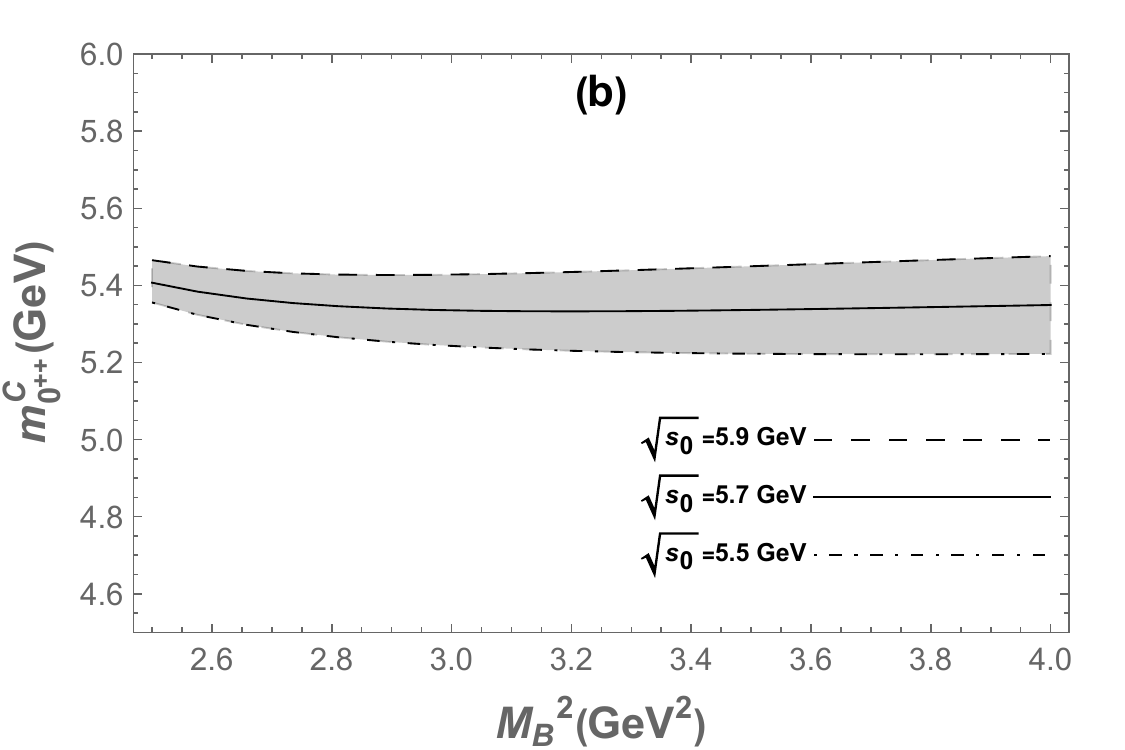}
\caption{Similar captions as in Fig.~\ref{fig0++B}, but for the current in Eq.~(\ref{current-0++C}).} \label{fig0++C}
\end{figure}

\begin{figure}
\includegraphics[width=6.8cm]{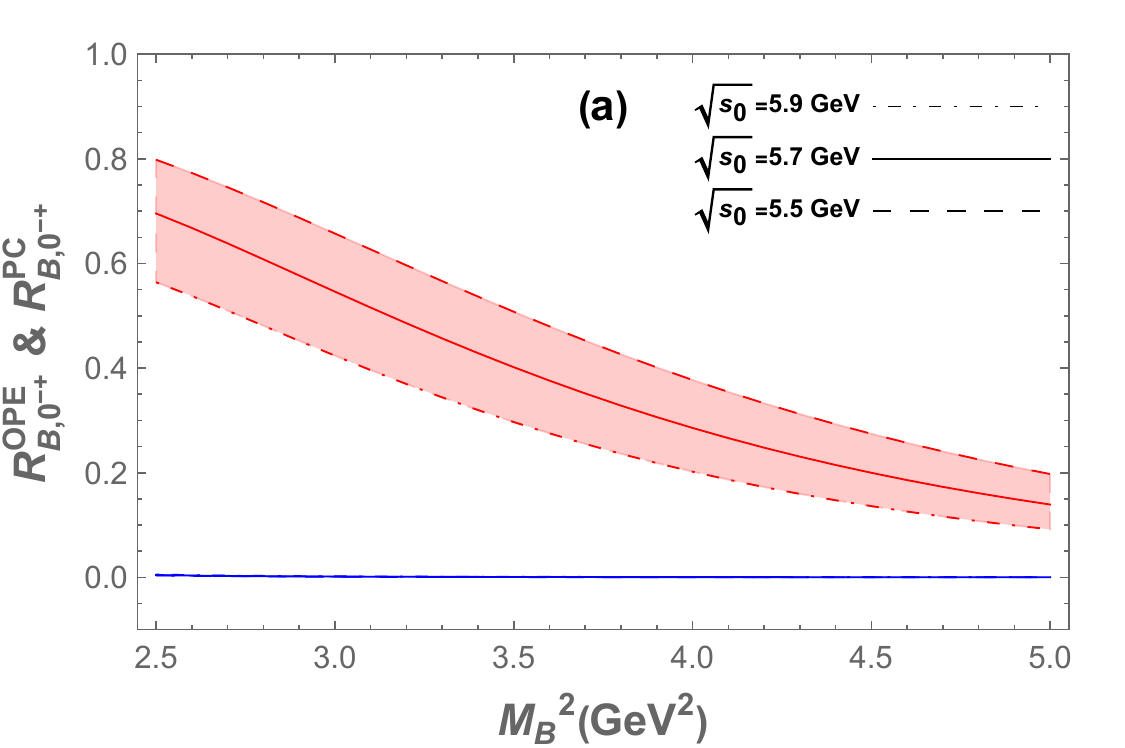}
\includegraphics[width=6.8cm]{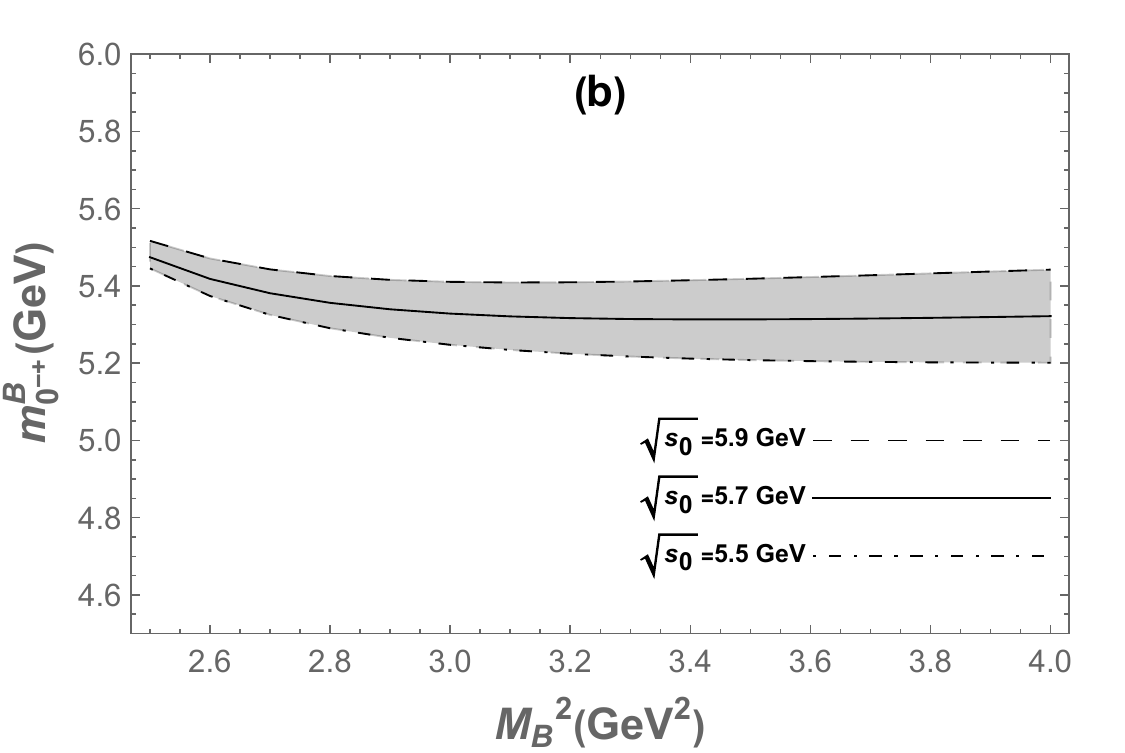}
\caption{Similar captions as in Fig.~\ref{fig0++B}, but for the current in Eq.~(\ref{current-0-+B}).} \label{fig0-+B}
\end{figure}

\begin{figure}
\includegraphics[width=6.8cm]{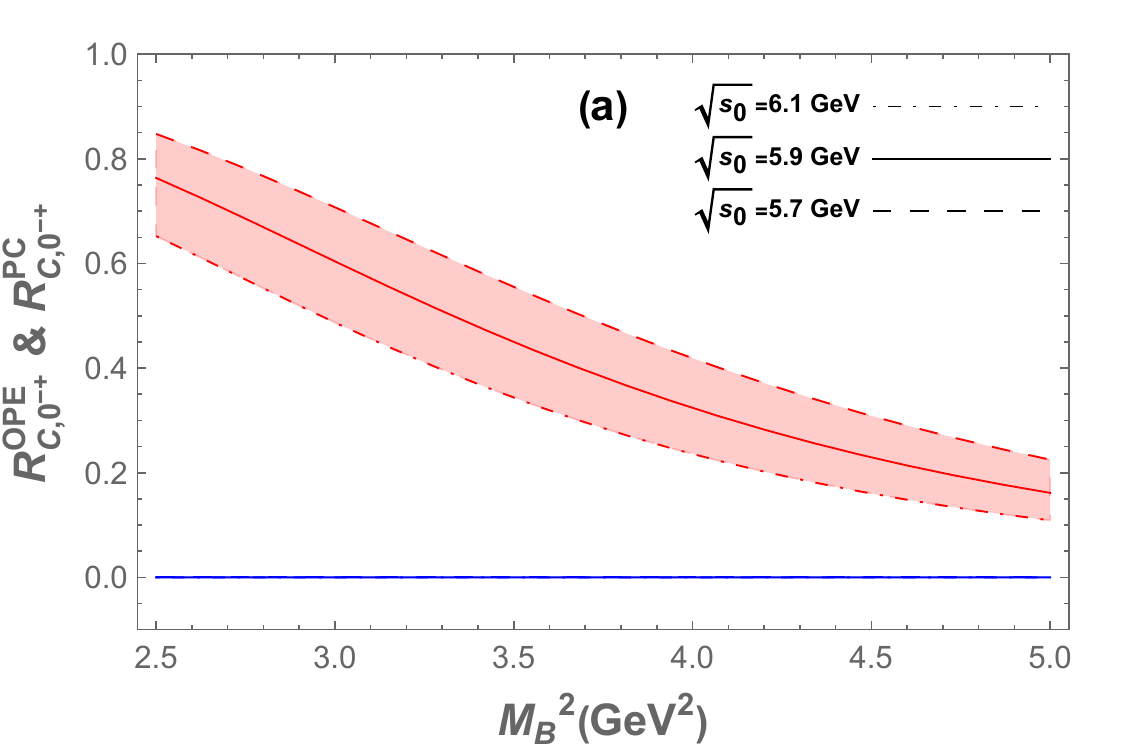}
\includegraphics[width=6.8cm]{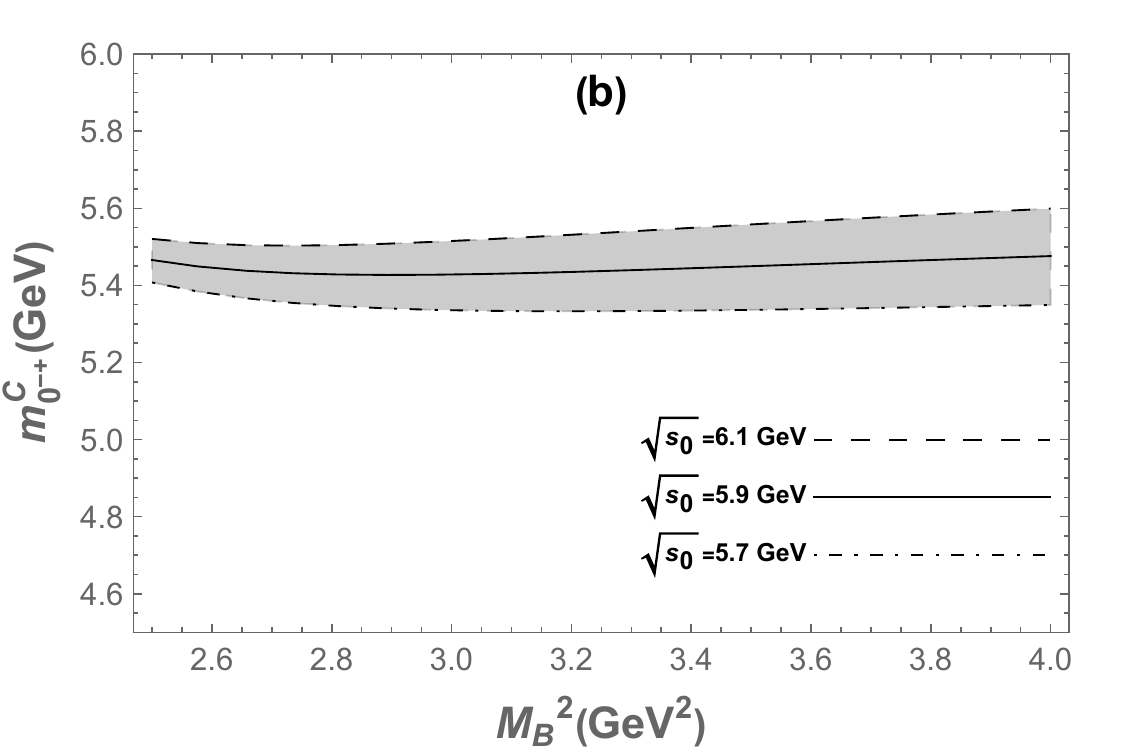}
\caption{Similar captions as in Fig.~\ref{fig0++B}, but for the current in Eq.~(\ref{current-0-+C}).} \label{fig0-+C}
\end{figure}

\begin{figure}
\includegraphics[width=6.8cm]{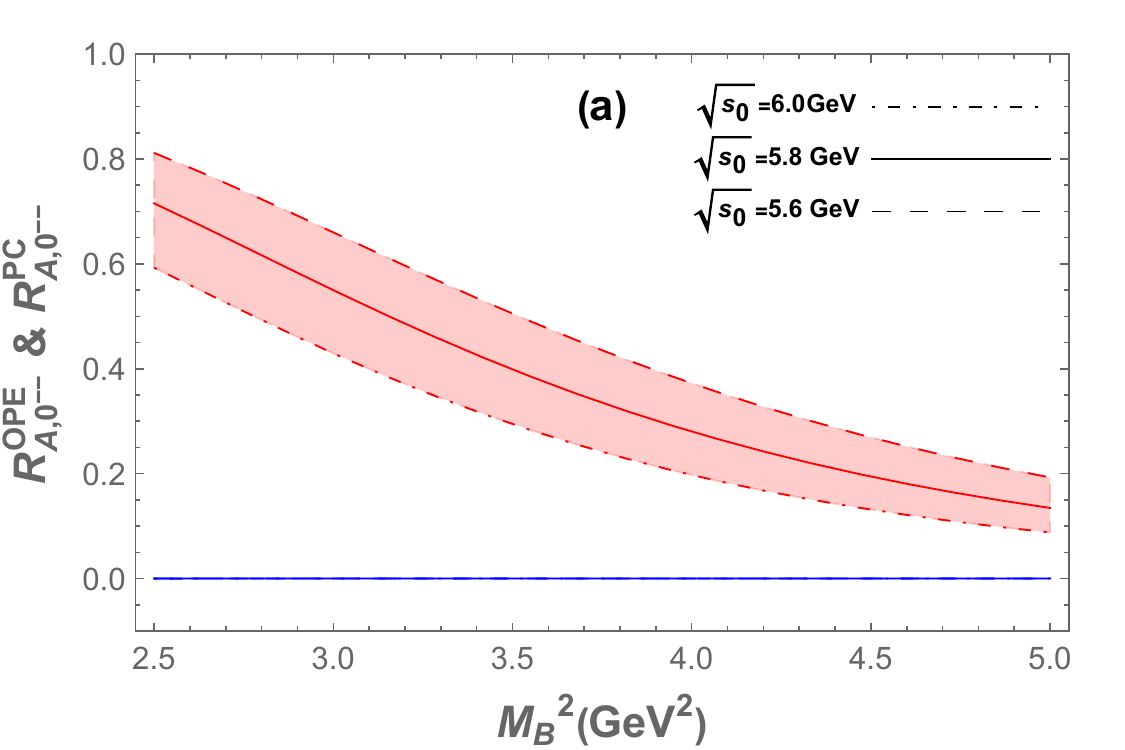}
\includegraphics[width=6.8cm]{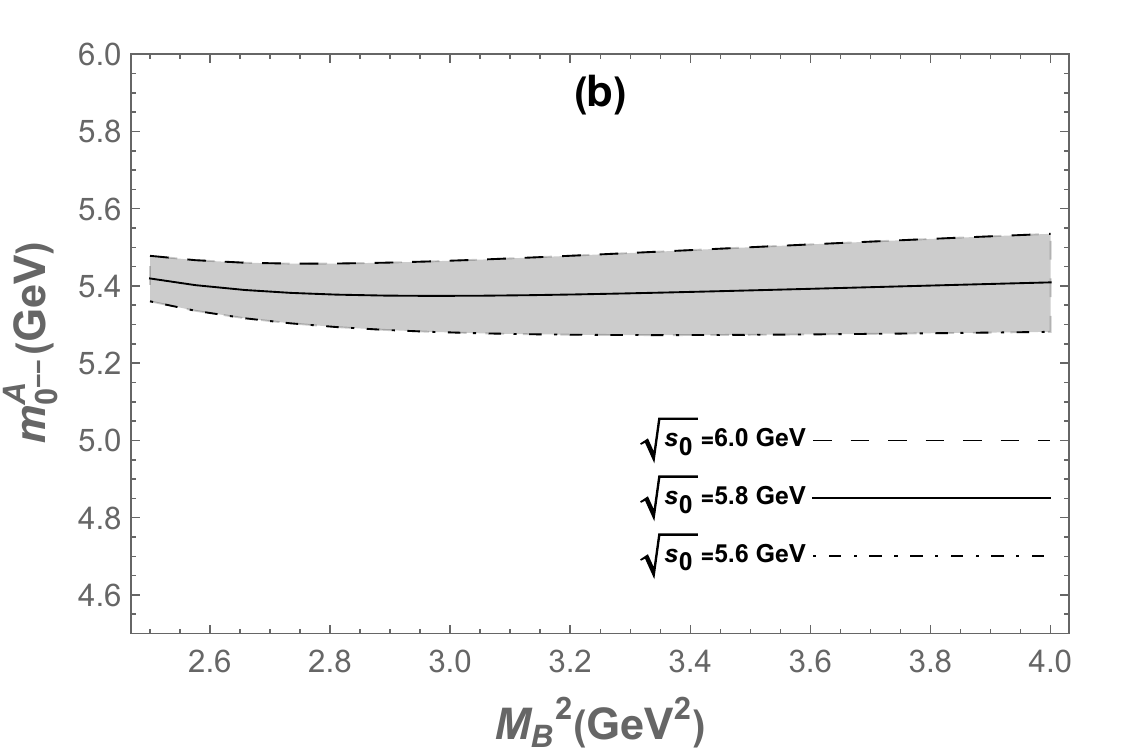}
\caption{Similar captions as in Fig.~\ref{fig0++B}, but for the current in Eq.~(\ref{current-0--A})} \label{fig0--A}
\end{figure}

\begin{figure}
\includegraphics[width=6.8cm]{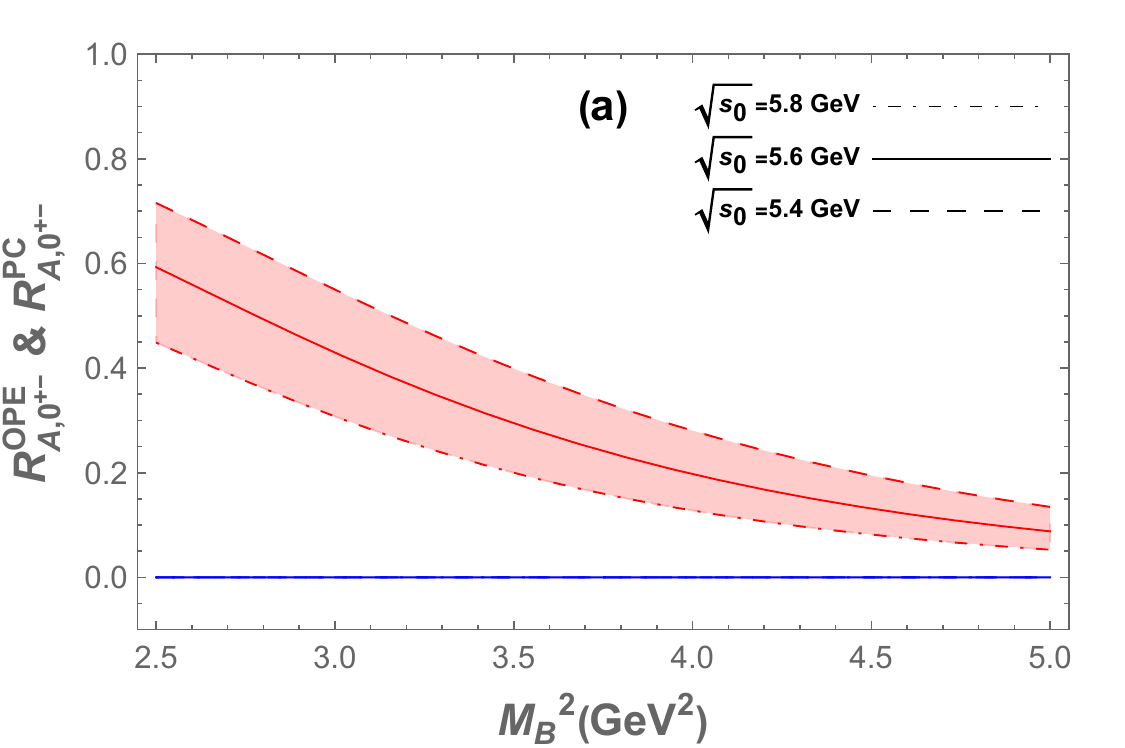}
\includegraphics[width=6.8cm]{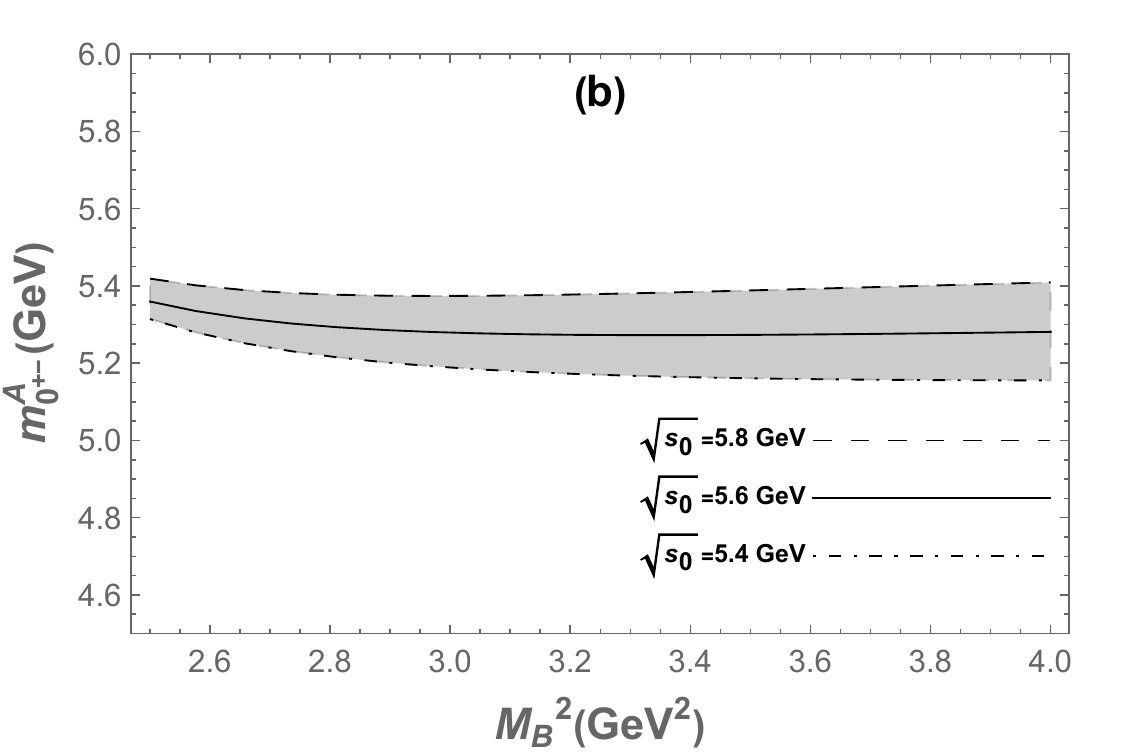}
\caption{Similar captions as in Fig.~\ref{fig0++B}, but for the current in Eq.~(\ref{current-0+-A}).} \label{fig0+-A}
\end{figure}

\begin{figure}
\includegraphics[width=6.8cm]{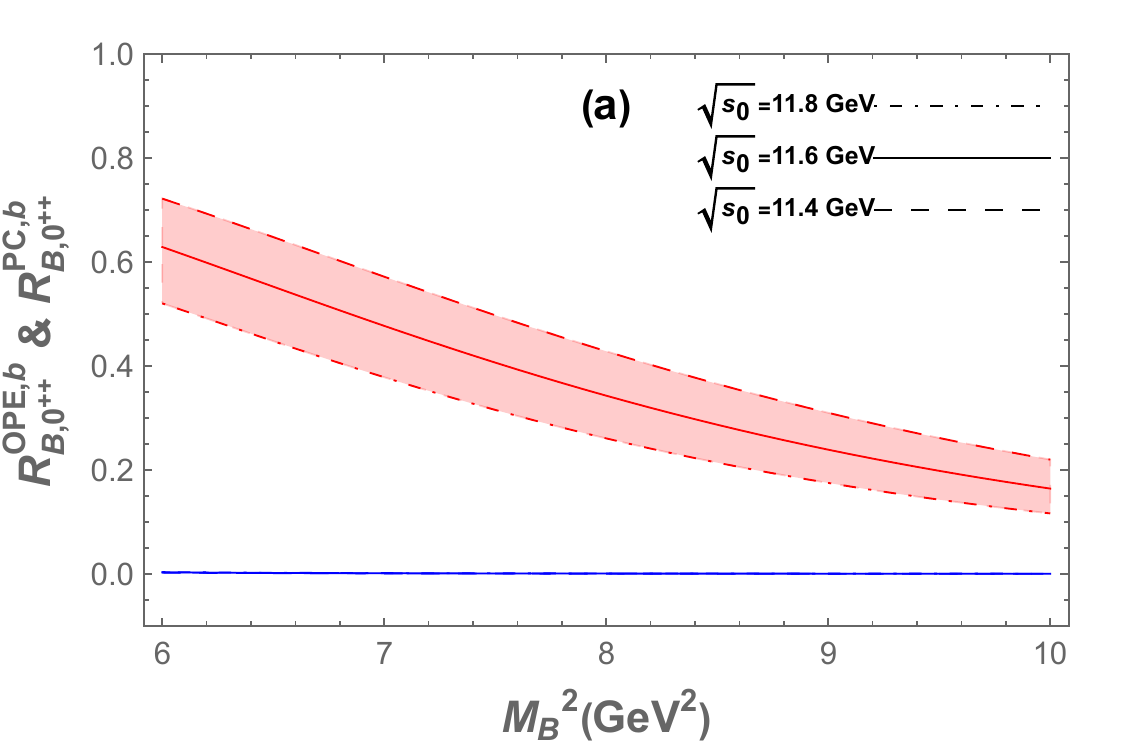}
\includegraphics[width=6.8cm]{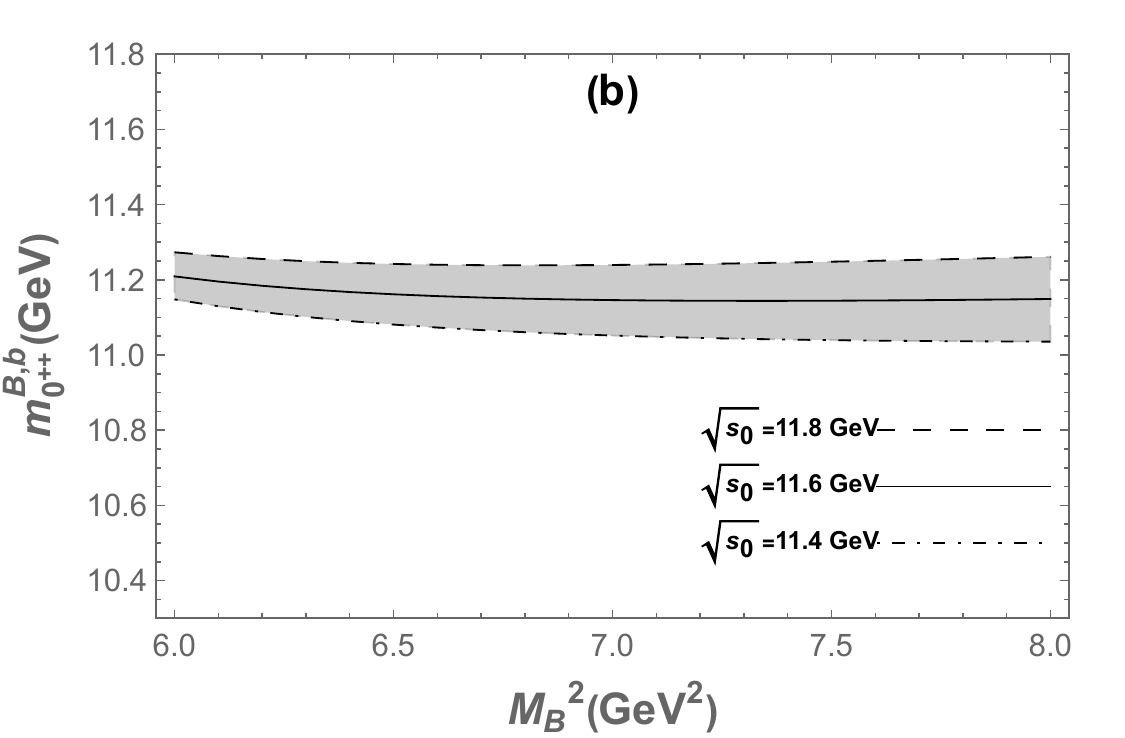}
\caption{Similar captions as in Fig.~\ref{fig0++B}, but for the $b$-sector and for the current in Eq.~(\ref{current-0++B}).} \label{fig0++Bb}
\end{figure}

\begin{figure}
\includegraphics[width=6.8cm]{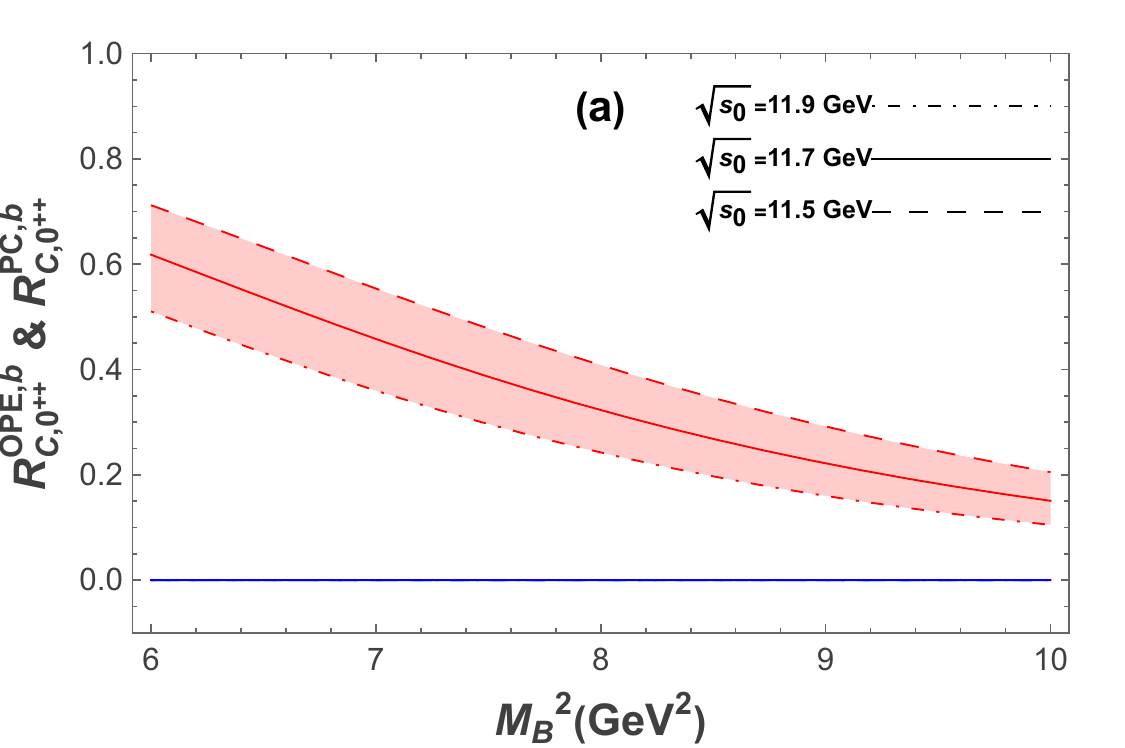}
\includegraphics[width=6.8cm]{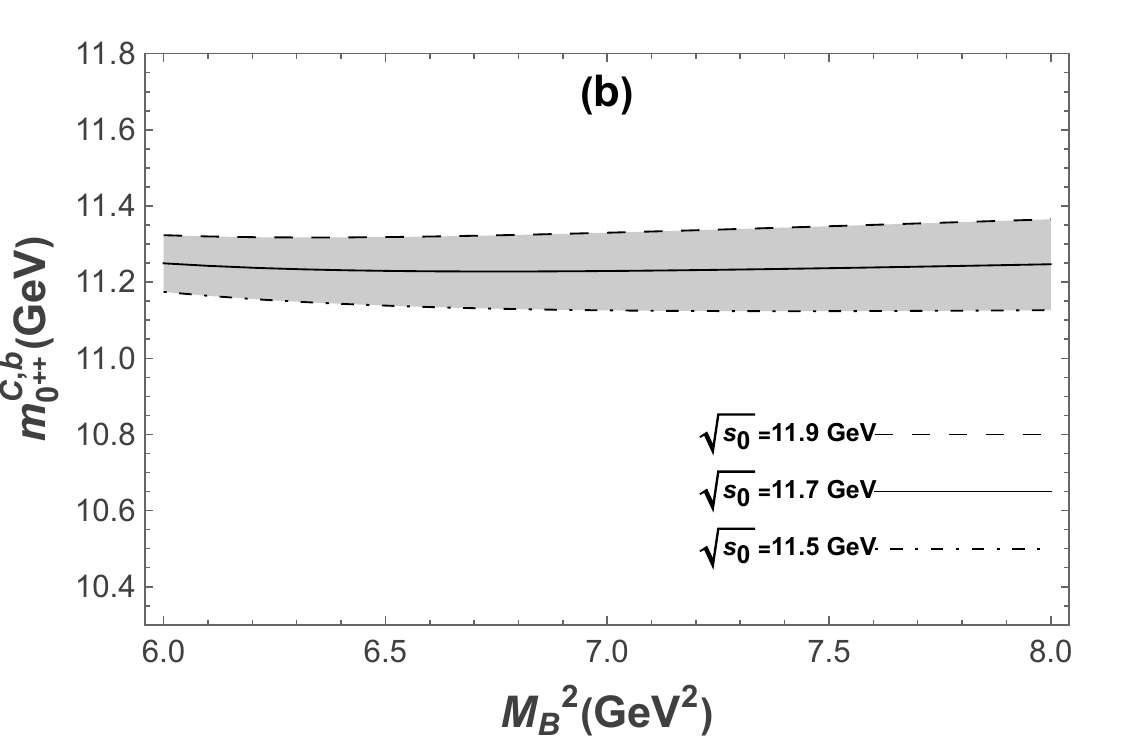}
\caption{Similar captions as in Fig.~\ref{fig0++B}, but for the $b$-sector and for the current in Eq.~(\ref{current-0++C}).} \label{fig0++Cb}
\end{figure}

\begin{figure}
\includegraphics[width=6.8cm]{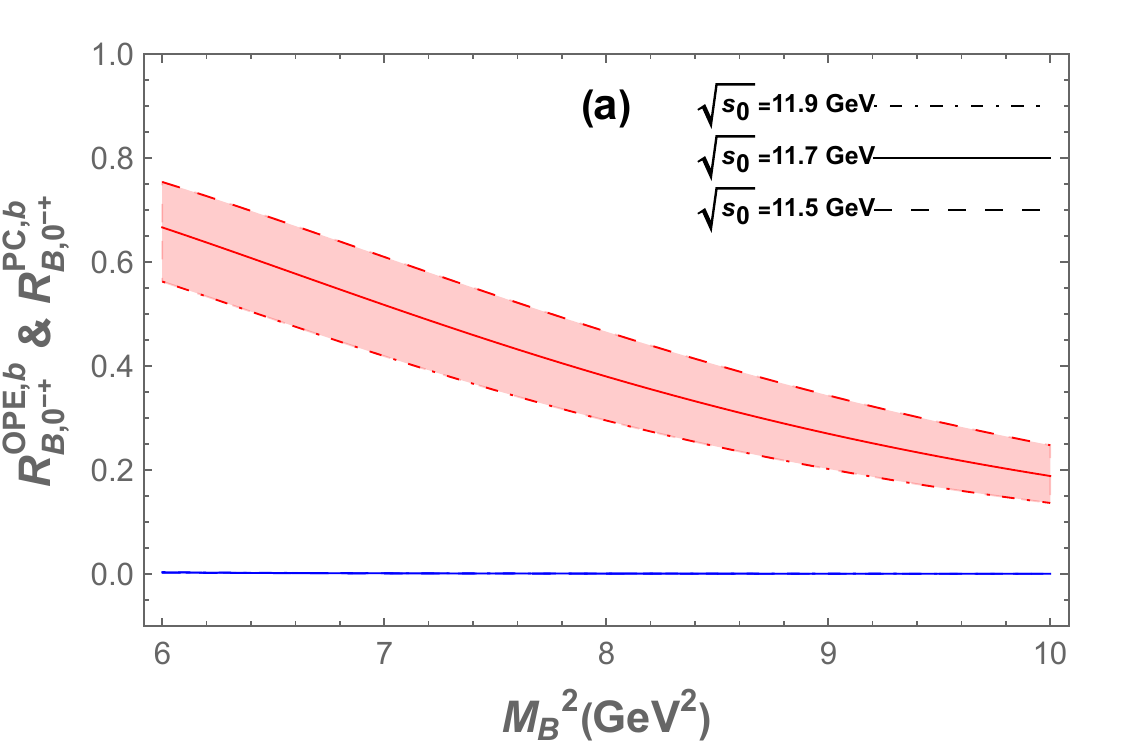}
\includegraphics[width=6.8cm]{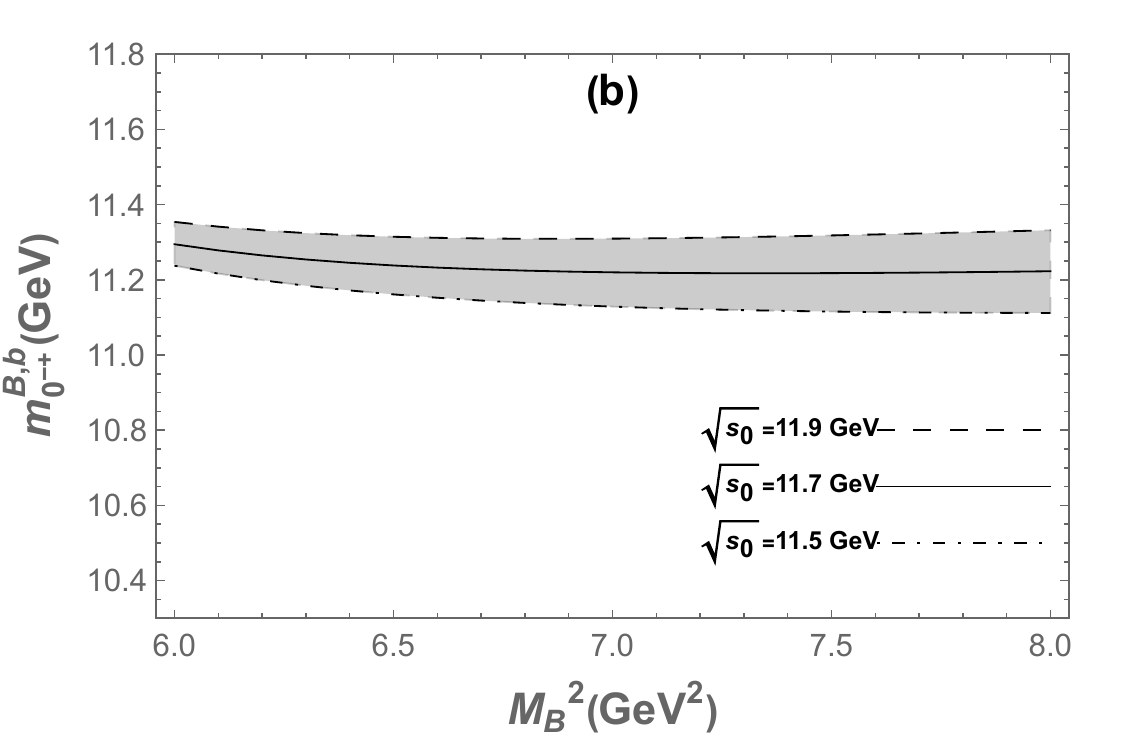}
\caption{Similar captions as in Fig.~\ref{fig0++B}, but for the $b$-sector and for the current in Eq.~(\ref{current-0-+B}).} \label{fig0-+Bb}
\end{figure}

\begin{figure}
\includegraphics[width=6.8cm]{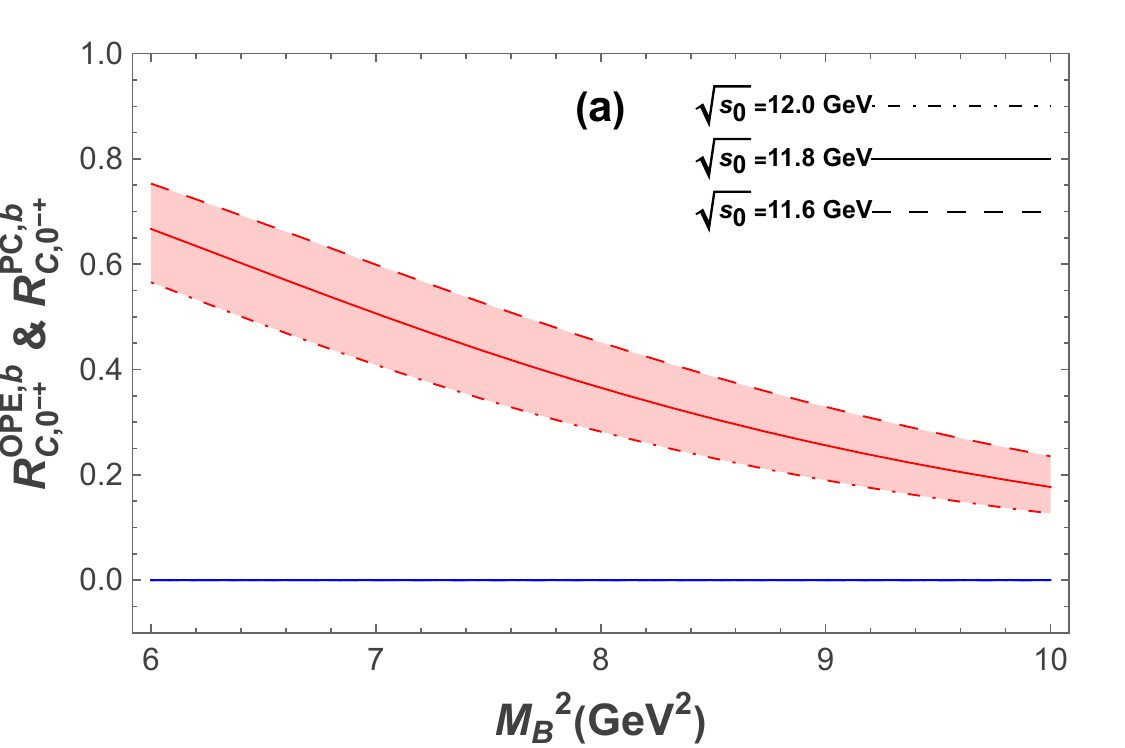}
\includegraphics[width=6.8cm]{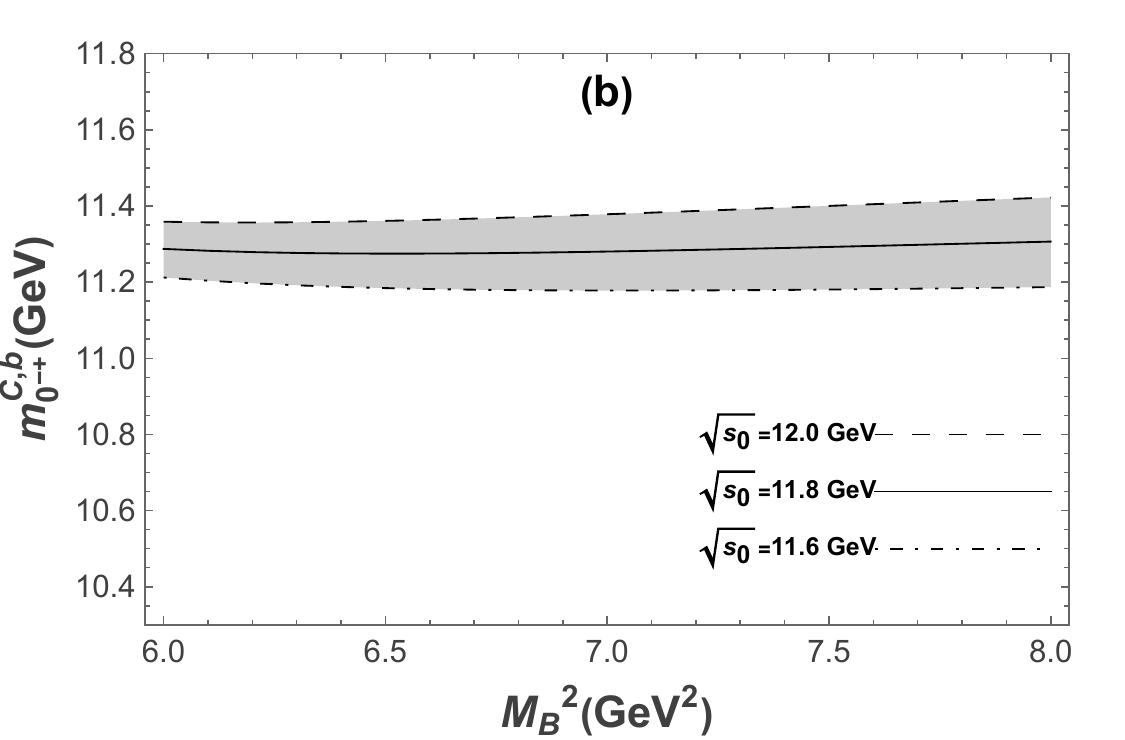}
\caption{ Similar captions as in Fig.~\ref{fig0++B}, but for the $b$-sector and for the current in Eq.~(\ref{current-0-+C}).} \label{fig0-+Cb}
\end{figure}

\begin{figure}
\includegraphics[width=6.8cm]{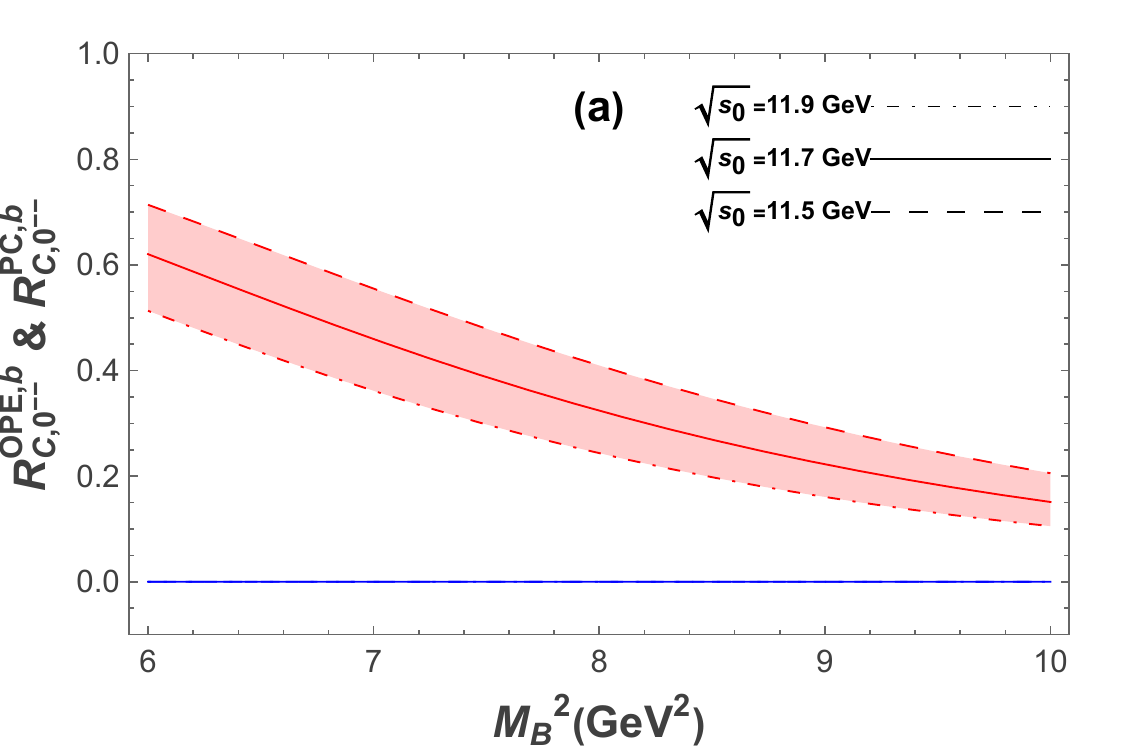}
\includegraphics[width=6.8cm]{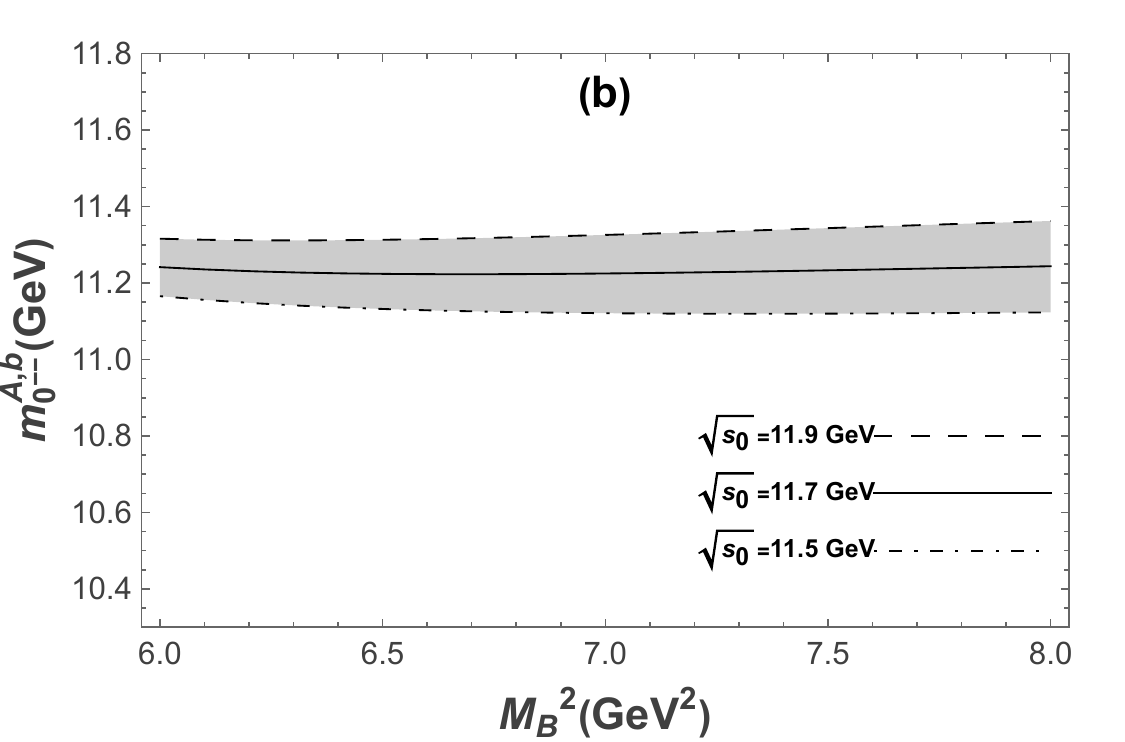}
\caption{Similar captions as in Fig.~\ref{fig0++B}, but for the $b$-sector and for the current in Eq.~(\ref{current-0--A}).} \label{fig0--Ab}
\end{figure}

\begin{figure}
\includegraphics[width=6.8cm]{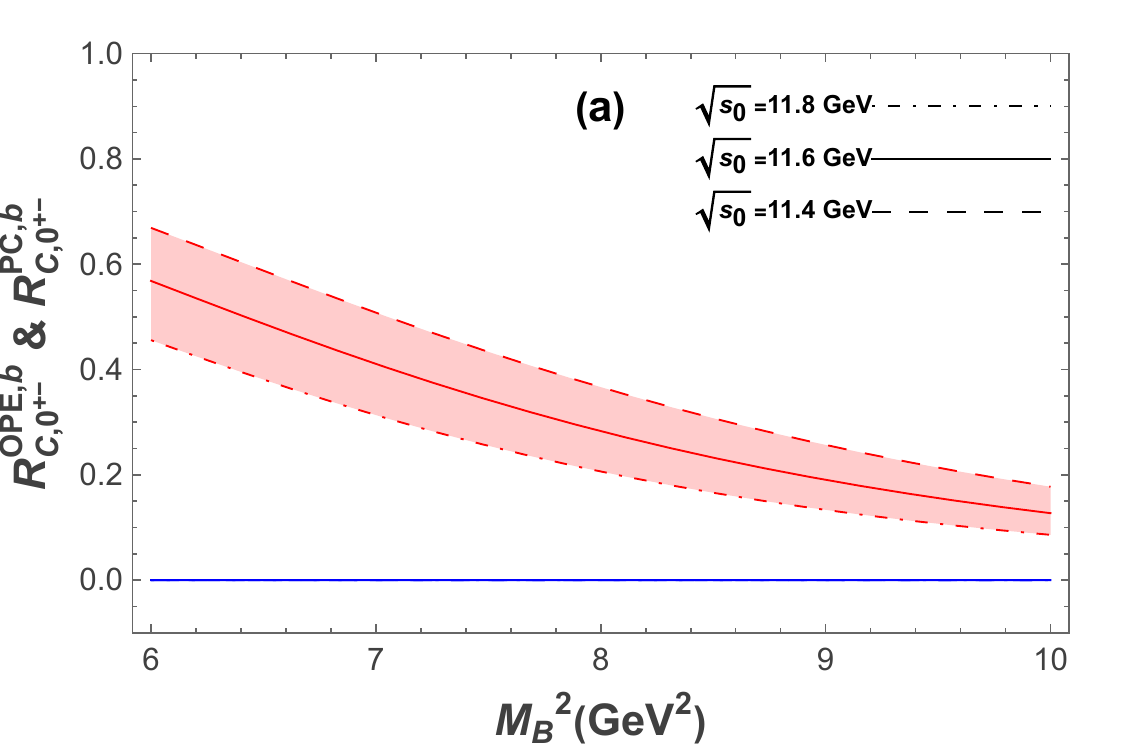}
\includegraphics[width=6.8cm]{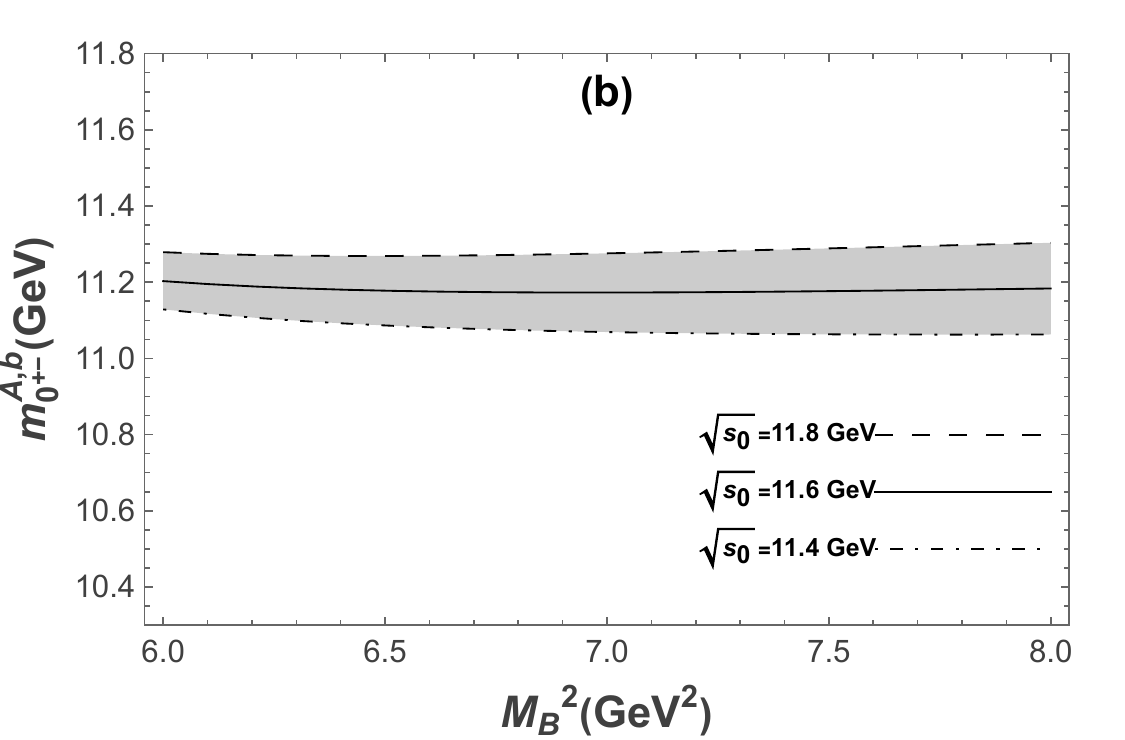}
\caption{ Similar captions as in Fig.~\ref{fig0++B}, but for the $b$-sector and for the current in Eq.~(\ref{current-0+-A}).} \label{fig0+-Ab}
\end{figure}

\end{widetext}

\end{document}